\documentclass[prb,preprint, floatfix]{revtex4-1} 
\usepackage{amsmath} 
\DeclareMathOperator{\Tr}{Tr}
\usepackage{amsfonts} 
\usepackage{graphicx} 
\usepackage{braket}
\usepackage[dvipsnames]{xcolor}
\usepackage{hyperref}
\begin{document}

\title{Geometric Visualizations of \\Quantum Mixed States and Density Matrices}
\author{Athanasios Kostikas, Yaroslav Valchyshen, and Paul Cadden-Zimansky}
\email{paulcz@bard.edu} 
\affiliation{Physics Program, Bard College, Annandale-on-Hudson, NY 12504}
\date{\today}

\begin{abstract} 
This paper presents an introduction to geometric representations of quantum states in which each distinct quantum state, pure and mixed, corresponds to a unique point in a Euclidean space.  Beginning with a review of some underappreciated properties of the most commonly used geometric representation, the Bloch sphere visualization of qubit states, we show how concepts, algorithms, and spatial relations viewable on this geometric representation can be extended to representations of qudit states of any finite quantum dimension $d$ and on to the infinite-dimensional limit.  A primary goal of the work is helping the reader develop a visual intuition of these spaces, which can complement the understanding of the algebraic formalism of quantum mechanics for learners, teachers, and researchers at any level.   Particular emphasis is given both to understanding states in a basis-independent way and to understanding how probability amplitudes and density matrix elements used to algebraically represent states in a particular basis correspond to line segments and angles in the geometric representations.  In addition to providing visualizations for such concepts as superpositions, mixtures, decoherence, and measurement, we demonstrate how the representations can be used to substitute simple geometrical calculations for more cumbersome linear algebra ones, which may be of particular use in introducing mixed states and density matrices to beginning quantum students at an early stage.  The work concludes with the geometrical interpretation of some commonly used metrics such as the purity of states and their relation to real, Euclidean vectors in the infinite-dimensional limit of the space, which contains all lower-dimensional qudit spaces as subspaces.
\end{abstract}

\maketitle 

\section{Introduction} 

In the teaching of classical mechanics, it is customary to begin by presenting simple descriptions of idealized situations -- think of frictionless planes and projectile motion with no air resistance -- in order to first introduce correspondingly simple mathematical descriptions and rules that ignore the complicating factors of dissipation and entropy.  In doing so, there is little risk of students confusing the idealization for the more complex reality of how classical objects move in their environment, as they have many years of experience with friction and air resistance.  The student tacitly or explicitly understands that the formal picture they initially encounter is only a starting point that can be amended later to account for richer, more complex, and more accurate descriptions of reality.

In the teaching of quantum mechanics, the initial presentation of the descriptive formalism is invariably idealized in a similar fashion.  Most quantum textbooks and courses focus almost entirely on pure-state descriptions of objects,\cite{griffiths, buzzell} represented by wavefunctions or complex vectors, that model only entropy-free, isolated systems subject to no entangling interactions with their environment, save for a nebulously defined measurement process.  However, in the quantum case, the more complex formal structure of the material and students' lack of prior physical intuition greatly obscure their awareness that what they are learning is a very limited description only applicable to the \textit {zero-entropy fantasy world} previously encountered in classical mechanics.  To cite one concrete example, the simplest quantum object to describe is a qubit state, such as the spin state of an electron or the polarization state of a photon.  Ask the following question to a student who has spent time learning quantum mechanics:  ``Pick any electron in your body or any photon that is hitting you; what is the quantum description of the electron's spin or the photon's polarization?''  The degree to which the student does not understand that the needed qubit description is not a pure state represented by a complex vector, but a mixed state represented by a density matrix, is the degree to which they have a limited ability to characterize the quantum nature of the world around them and the degree to which they are in danger of mistaking the zero-entropy fantasy world of the pure state formalism for a faithful quantum model of reality.

This danger, coupled with the increasing importance of mixed states and density matrices in quantum information research and other subfields of physics, creates a motivation to introduce and teach these concepts as early as possible to physics students and as widely as possible to the growing number of non-physics students enrolling in quantum information courses.  The manifest challenge of early introduction is the fact that the syntactical algorithms of the mixed state formalism are more complicated than the limited, pure state formalism. Hence, it makes sense to first acclimate students to the pure state algorithms and concepts and only then, time permitting, introduce mixed states.  To circumvent this challenge, in this paper we present an alternate and complementary visual and geometric approach to understanding quantum states that allows for early introduction and intuition-building around mixed states.

This work is a guide to geometrically visualizing quantum states, pure and mixed, within Euclidean spaces:  each state corresponds to a point in the space, a \textit {statepoint}, related to other states' points by meaningful distances governed by familiar Euclidean geometry, which allow one a direct understanding of both measurement probabilities for any measurement and the complex number \textit{coordinates} used in column vector or density matrix representations of the state in any basis.  This \textit{statespace} representation technique can be used for quantum states of any finite-dimensional basis with some results extendable to the infinite-dimensional and continuum limits.  In almost all cases, the full statespace representation exists in dimensions higher than the 3-dimensional spaces we can fully intuit, but, as we show, the challenge of picturing these spaces can be circumvented by considering visualizable subspaces within them and using precise analogies with lower-dimensional statespaces that are intuitive.  The use of such analogies is already invoked in the learning of quantum mechanics -- think of how the visualizable ideas of normalized or orthogonal real space 2-dimensional and 3-dimensional vectors are useful to have in mind when encountering the more generalized notions of orthogonal or normalized state vectors or wavefunctions in the quantum formalism.  Indeed, the notion of vector as arrow is helpful in understanding it as a mathematical object independent of the many possible coordinate representations it can be specified in.  However, most students, and perhaps most physicists, have no corresponding visual intuition for linear maps, tensors, and mixed state density operators and their matrix representations.  A central goal of this work is to present visual pictures allowing one to understand mixed states independently of any matrix representation -- a point in a statespace -- while also giving geometrical interpretations to the numerical entries of density matrix representations of these states in any chosen basis.  We end by showing a one-to-one correspondence between every finite dimensional state, pure and mixed, and a real vector drawn from a suitable origin to its unique statepoint.

The geometric representation techniques presented have the property that they can be used at multiple levels, introducing the basic concepts of quantum states before teaching any formalism (indeed one can use them to introduce the concept of mixed states {\it before} pure-state concepts such as superposition).  For students learning the quantum formalism, the geometric representations provide a useful accompaniment that can help motivate and teach mixed-state descriptions both qualitatively and quantitatively. Important concepts such as decoherence, often omitted from undergraduate courses due to their formal complexity, can be introduced using them.  While the Bloch sphere statespace of a 2-level qubit has a long history\cite{stokes,poincare,fano,mcmaster} and wide usage in quantum teaching and understanding,\cite{nielsen,schumacher} several useful and intuitive aspects of how to read column-vector and density-matrix entries off of it remain largely absent from textbooks.  Research into higher-dimensional statespaces, such as 3-level ``qutrit'' states\cite{tabia,goyal,eltschka} and general $d$-level ``qudit'' states\cite{kimura,bertlmann} has increased in the past few decades, paralleling the growth of the field of quantum information, but to the authors' knowledge this research has only been explained in detail at the textbook level in a single, graduate-level work.\cite{bengtsson} Our work aims for an undergraduate-level explication of the framework and utility of the geometric representations, but is likely also of use to those at the graduate and research levels unfamiliar with them, since they provide a complementary tool to the algebraic formalism for exploring the possibilities of quantum state evolution and in some cases provide methods of improving computational efficiencies.

Our presentation begins by reviewing some underappreciated properties of the Bloch-sphere geometric representation for the simplest quantum states: the $d=2$ dimensional basis space of a qubit.  In doing so we emphasize multiple properties of this statespace that extend to higher-dimensional cases and show how these manifest, using examples in the $d=3$ and $d=4$ cases, in order to understand general results that apply in any dimension.  Throughout the main body of the paper, we will focus on building conceptual intuitions for the geometry of these statespaces and quote some key results; proofs of quoted results are included in the appendices using derivations accessible to undergraduate physics students.  
\section{Introducing Mixed States}
\begin{figure}[h!]
\centering
\includegraphics[scale=.48]{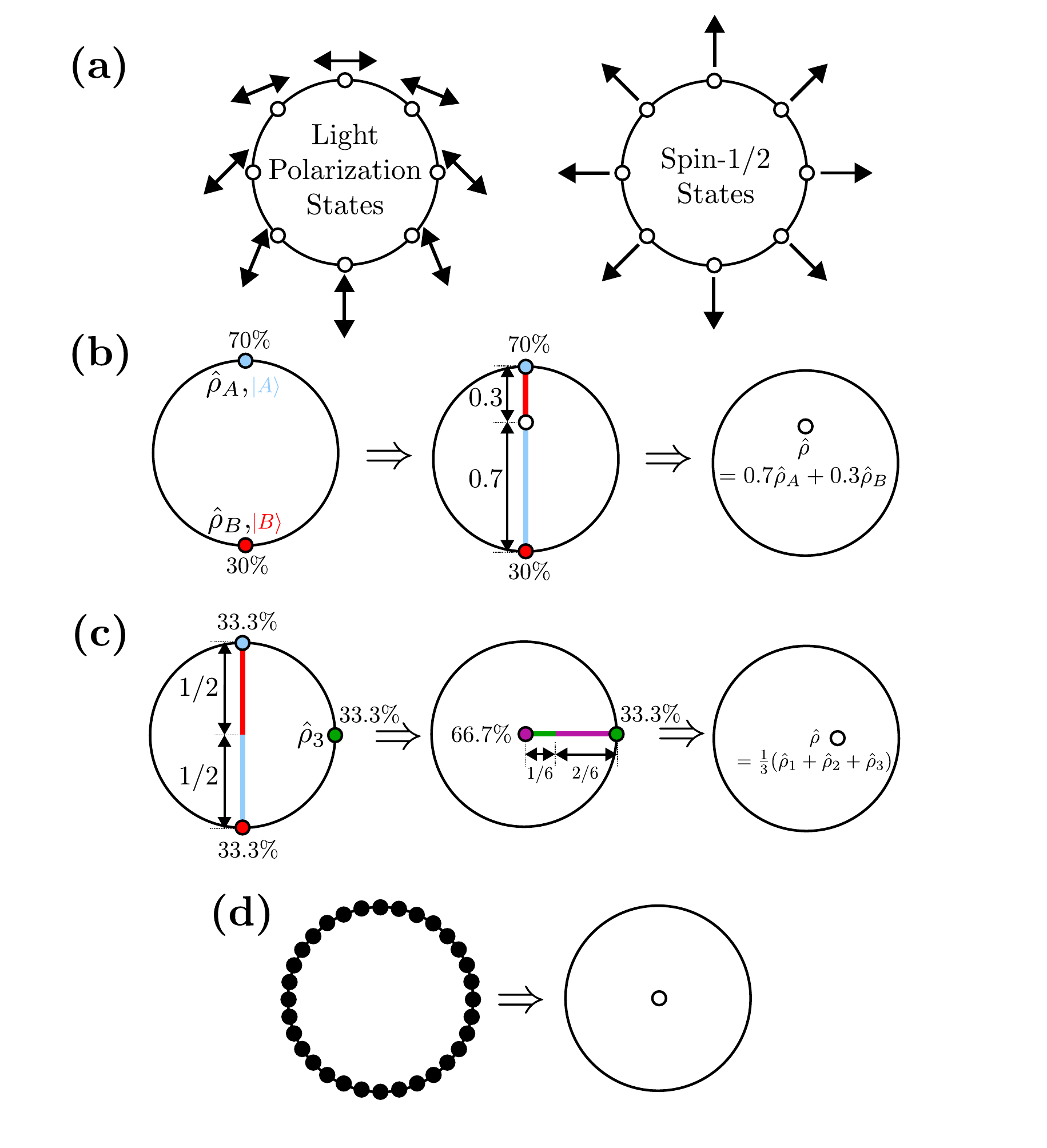}
\caption{(a) A selection of pure qubit statepoints on a Bloch circle cross section of a Bloch sphere corresponding to different linear light polarization states (left) and different spin-1/2 states (right). (b) The statepoint representing a 70\% chance of an up state and a 30\% chance of a down state is found by dividing the line connecting these two points in segments corresponding to the probabilities.  (c) In general, statepoints can be found using a center-of-mass algorithm with probabilities replacing masses, such as is shown for the state with equal likelihood of being an up, down, or right state. (d) By this algorithm, it is clear that the statepoint for the state with equal likelihood of all pure states is the center of the circle, corresponding to the maximally mixed qubit state.}
\label{figBlochCircle}
\end{figure}

The Bloch Sphere is the most widely used statespace with each possible statepoint on or in the sphere corresponding 1-to-1 to a qubit quantum state.  As mentioned above, common qubit examples are the magnetic moment orientation of spin-1/2 particles or photon polarizations.  For each definite spin direction or light polarization there is a corresponding statepoint on a sphere.  Fig.~\ref{figBlochCircle}(a) shows circular cross sections of this sphere for plotting either the spin state as magnetic moment arrows or linear polarization states of light.  (The full sphere includes points for all 3-dimensional spin directions, including elliptical and circular light polarizations.)  Each of these statepoints on the boundary of the sphere corresponds to a unique pure state denoted by a density operator such as $\hat{\rho_A}$ or ket $\ket{A}$. Any two antipodal statepoints on opposite sides of the sphere boundary are orthogonal states.

 Now consider the following: you are given a qubit that has a probability $p_A$ (e.g. a 70\% chance) of being in one state $\hat{\rho}_A$ and a $p_B=1-p_A$ chance (e.g. 30\%) of being in the opposite, orthogonal state $\hat{\rho}_B$.  Note that these probabilities can be viewed -- emphasis on {\it can} -- as classical, epistemic probabilities of ignorance that are more familiar to students and thus more intuitive than the ontic probabilities that arise in quantum measurements of pure-state superpositions.  Following Fig.~\ref{figBlochCircle}(b), to find the statepoint for this \textit {mixed state}, stipulate that the sphere has a radius 1/2, draw the unit diameter line connecting the $\hat{\rho}_A$ and $\hat{\rho}_B$ statepoints, and locate the point on this diameter that cuts it in a 70/30 ratio and is closer to the more likely state.  This point corresponds to the statepoint of this mixed state, algebraically calculated as $\hat{\rho}=0.7\hat{\rho}_A+0.3\hat{\rho}_B$.

The specification of a mixed state is not limited to two probabilities of two orthogonal states:  Let $\hat{\rho}_1,\hat{\rho}_2,\hat{\rho}_3,\ldots\hat{\rho}_n$, be any number of possible qubit states, located at statepoint coordinates $\bf{r}_1,\bf{r}_2,\bf{r}_3,\ldots\bf{r}_n$ with respect to any choice of Euclidean coordinates.  The mixed state with a $p_1$ probability of being in state $\hat{\rho}_1$, a $p_2$ probability of being in state $\hat{\rho}_2$, etc., is formally calculated by $\hat{\rho}=\sum_{i=1}^{n} p_i\hat{\rho}_i$ corresponding to a statepoint located at ${\bf r}=\sum_{i=1}^{n}p_i\bf{r}_i$ (see Appendix \ref{App_QubitBary} for a proof).  This ``barycentric'' algorithm is identical to the center-of-mass algorithm used in introductory physics with the probability weightings replacing the mass weightings and the division by the sum of all probabilities being unnecessary as long as this sum is normalized to one.  This similarity makes the location of statepoints in the quantum statespace intuitive for anyone familiar with the center-of-mass concept.  The 70/30 example above shows the case of a center-of-probability statepoint being located on the line between the two statepoints being ``weighted'' by their probabilities, the same as the center of mass between two masses in this ratio.

While the formula gives an exact location that depends on a choice of coordinate system, one can alternately use geometric techniques to locate the statepoint corresponding to a state without reference to any coordinate system.  Consider the statepoint of an object that has a 1/3 chance of being in each of the three states shown in Fig.~\ref{figBlochCircle}(c), for example a spin pointing either up, down, or to the right with equal probability.  One can first find the center of probability of the two opposite points to be the center of the circle and then weight this combined $1/3+1/3 = 2/3$ probability against the remaining $1/3$ probability of being on the edge.  Connecting these two points and dividing it in a $2/3$ to $1/3$ ratio with the division closer to the more likely point locates the statepoint of this state.  One last example that can be done with no calculation is the case when all pure states are equally likely (Fig.~\ref{figBlochCircle}(d)).  One's center-of-mass intuition locates this \textit {maximally mixed state} immediately as corresponding to the statepoint at the center of the circle.
\section{Qubit Projective Measurements and Decoherence Leaves}
 \begin{figure}[h!]
\centering
\includegraphics[scale=.46]{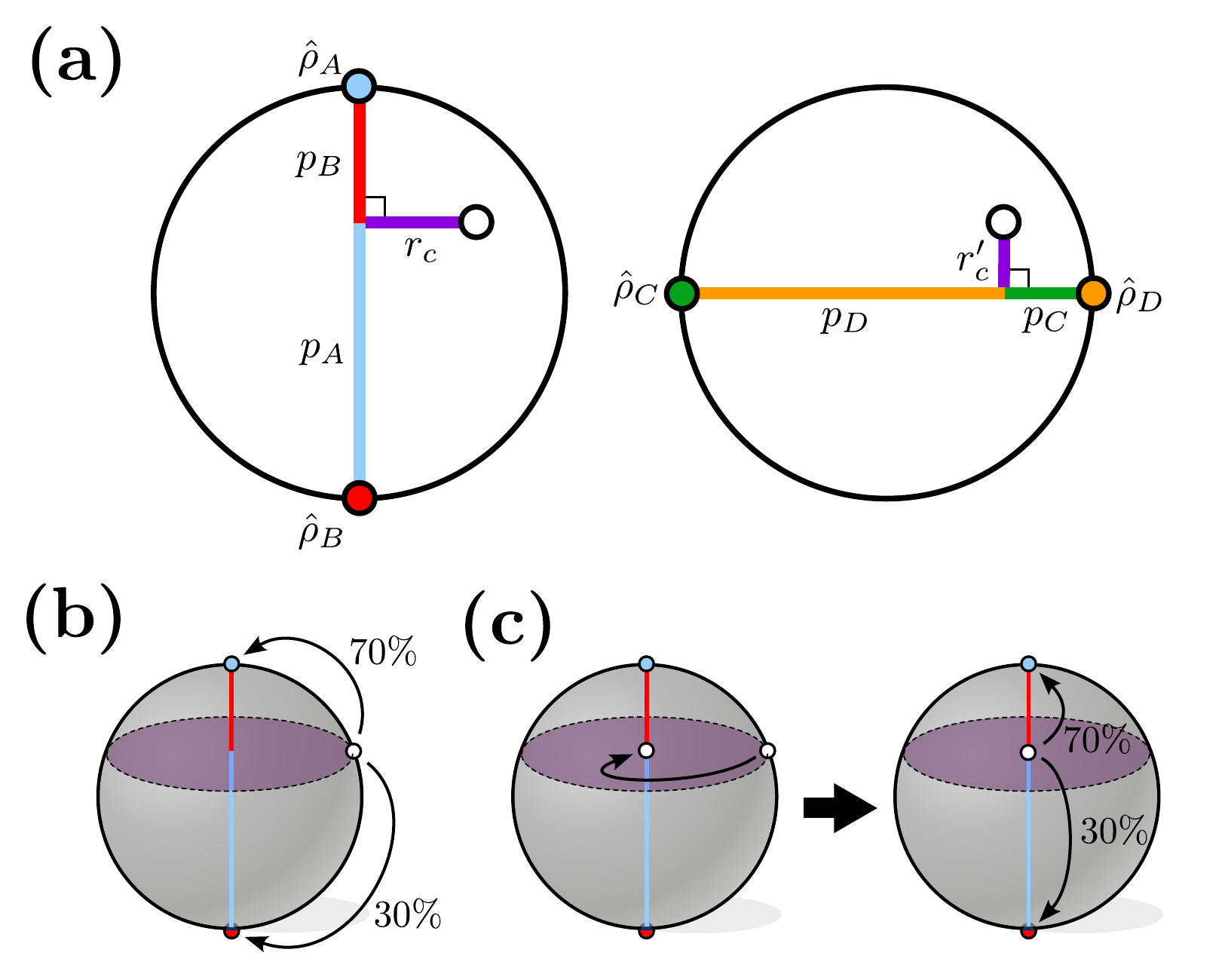}
\caption{(a) Geometrically, the probabilities of a projective quantum measurement on a qubit involve choosing a Bloch sphere diameter that terminates in two pure statepoints and drawing a perpendicular line from the statepoint being measured to this diameter.  The perpendicular projection cuts the diameter into two sections, equal in length to the probabilities of the measurement resulting in one of the two pure states.   (b)  A single indeterministic, irreversible process of measurement conventionally taught when only pure states are referenced:  a measured pure state evolves to one of two pure states determined by the choice of measurement diameter.  (c) The same measurement as a two-step process: an, in principle, deterministic decoherence process of the pure state evolving in the decoherence leaf disk to a mixed state along the measurement diameter, followed by an indeterministic outcome of the measured mixed state.}
\label{figMeasurement}
\end{figure}
The formalism of quantum mechanics is a method for calculating probabilities corresponding to frequencies of any possible measurement outcome on an object in any quantum state.  Here we limit ourselves to the projective value measurements that are simplest and usually the main focus for those initially learning quantum mechanics.  In the qubit case the possible choices of projective measurements correspond 1-to-1 to the possible choices of diameters of the Bloch sphere that terminate at two statepoints corresponding to two orthogonal states.  The two orthogonal states are the binary possible outcomes of this measurement.  The probability of each outcome is found geometrically by drawing a line from the statepoint of the state being measured that intersects the measurement axis perpendicularly (Fig.~\ref{figMeasurement}(a)).  The length of each part of the divided diameter gives the probabilities of each of the two outcomes, with the more probable outcome being the one closer to the statepoint measured.

As shown in Fig.~\ref{figMeasurement}(b) on a full Bloch sphere, it is conventional to present quantum measurements as a single process wherein an object in a state ``jumps'' to one of the possible measurement outcomes.  An alternate view, Fig.~\ref{figMeasurement}(c), has the diameter serve as a projective ``attractor'' of statepoints -- any statepoint evolves in a potentially deterministic, but irreversible way (insofar as the environment corresponding to the measurement apparatus is irreversible) in a plane perpendicular to the diameter to reach the statepoint on the diameter intersected by the plane.  This statepoint corresponds to a state which can be viewed as one on which a measurement has been made, but with epistemic probabilities as to the outcome.   The indeterministic, irreversible jump from this state to one of the two pure states occurs only once this outcome is epistemically determined.  The evolution of a statepoint through mixed states in the subspace of the sphere perpendicular to the axis is the subject of decoherence theory,\cite{zurek} a topic of considerable importance in the quantum theory of measurement, but one that is usually not taught at the undergraduate level due to its perceived complexity.  This geometrical picture allows one to introduce the concept and process without need for formal derivations.

As we shall see below in extending the geometry to higher dimensions, any given projective measurement and any given state define a subset of all the states having identical probabilities to the given one for that measurement; in the qubit case this ``decoherence leaf'' is this cross-sectional circular disk perpendicular to the given measurement diameter containing the corresponding statepoints of this set.  It is intuitively clear in the qubit case that for a given choice of measurement (diameter), the full statespace can be divided up into decoherence leaves (circular disks perpendicular to the diameter) such that every statepoint is in one and only one decoherence leaf.  Mathematically, this process of dividing up a space into subspaces is known as a \textit{foliation}, from whence the ``leaf'' terminology, with each choice of basis specifying a different foliation. Different foliations may be used to introduce the idea of quantum tomography,\cite{lvovsky} wherein repeatedly measuring systems in the same starting state using \textit {different} axes determines which decoherence leaf they are in for each measurement.  In the qubit case, measurements along two diameters narrows down the statepoint to be along the line of two intersecting decoherence leaf disks, and along three diameters narrows it down to the single statepoint of three intersecting disks.

\section{Finding Density-Matrix Representations Geometrically}
\begin{figure*}
	\centering
\includegraphics[scale=.72]{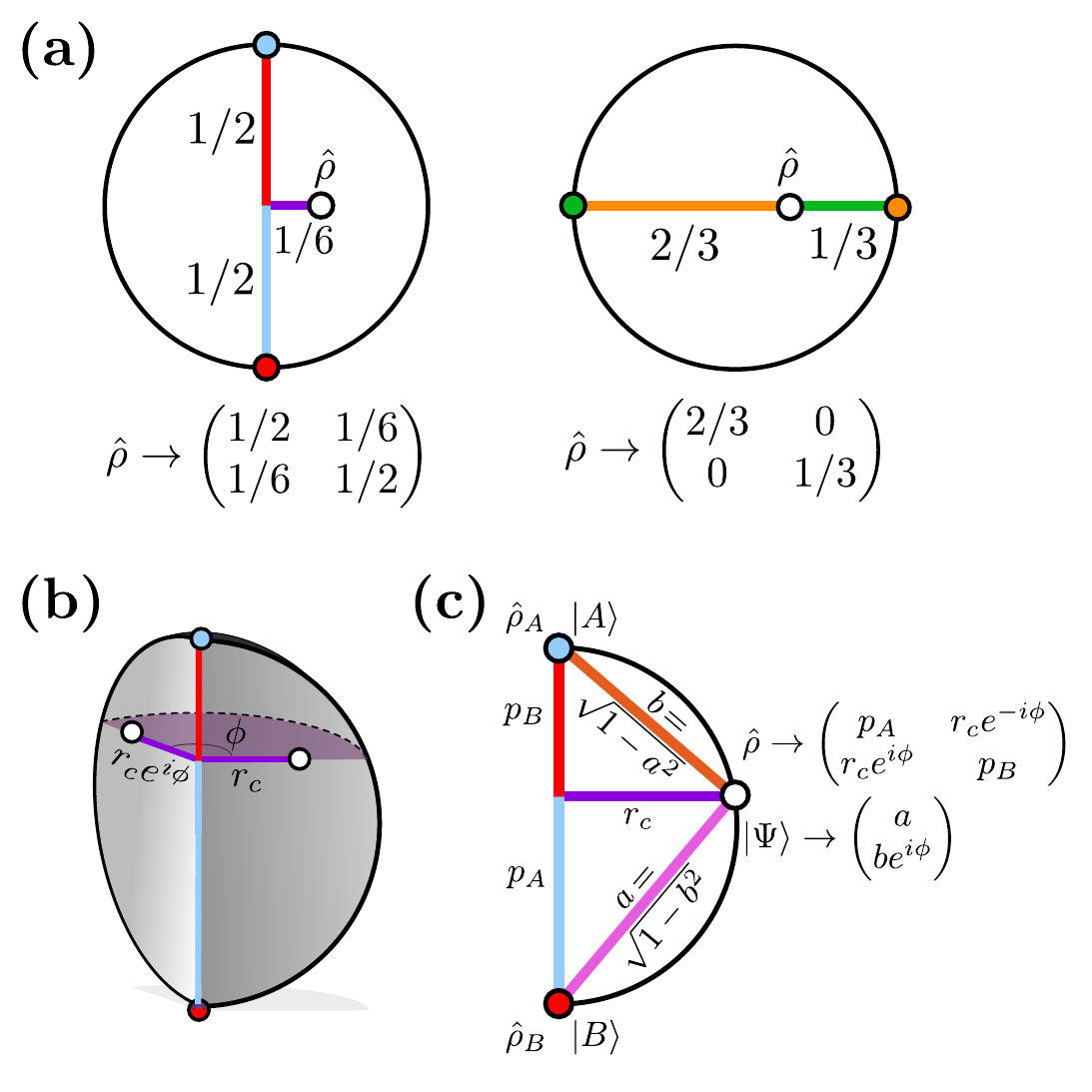}
\caption{(a) Representations of the state of Fig.~\ref{figBlochCircle}(c) in two different bases.  The matrix elements of the representations are equal to the lengths of the perpendicular from the statepoint to the basis diameter (off-diagonal) and the lengths of the segments of the diameter (diagonal elements). (b) A section of the full Bloch sphere made by rotating a semicircle about a possible measurement axis.  For a fixed set of measurement probabilities, the location of a statepoint within the ``decoherence leaf'' disk can be specified by a single complex number $r_ce^{i\phi}$ encoding its distance $r_c$ from and angle $\phi$ about the measurement diameter.  (c)  For any semicircular section of the Bloch sphere the $r_ce^{i\phi}$ decoherence leaf coordinate and its complex conjugate serve as the off-diagonal elements of the density matrix.  For pure statepoints, an alternate representation is a column vector, where the magnitudes of the entries $a$ and $b$ are equal to the lengths of the lines connecting the statepoint to the basis statepoints.}
\label{fig_representations}
\end{figure*}
The intuitive distinction between a point in space and a particular set of coordinates in an arbitrary coordinate system that locates that point is a useful framework for understanding the distinction between the basis-independent specification of a state $\hat\rho$ (statepoint) and its particular matrix representation (coordinates) in a basis $\rho$ (see Fig.~\ref{fig_representations}).  In the standard quantum formalism, the basis is specified by a complete set of orthogonal pure states, corresponding in the qubit case to any choice of Bloch sphere diameter terminating in two basis statepoints.  The coordinates of the mixed state $\hat\rho$ in this basis are recorded in a density matrix $\rho$.  The diagonal elements' probabilities correspond to the location along the basis diameter where the decoherence leaf containing the statepoint intersects it perpendicularly.  The off-diagonal elements locate the position of the statepoint within the decoherence leaf.  Since the qubit decoherence leaves are circular cross sections of the sphere, the coordinates of a statepoint within it can be located using polar coordinates:  $r_c$ gives the distance from the leaf's center on the basis diameter, $\phi$ gives it angular location about the basis diameter relative to an arbitrary zero angle set by basis convention (Fig.~\ref{fig_representations}(b)).  These real, polar coordinates are combined into a single complex number $r_ce^{i\phi}$ that is recorded in the off-diagonal element of the state's density matrix in the chosen basis.  The density matrix is hermitian, so that the other off-diagonal element is the complex conjugate $r_ce^{-i\phi}$ that redundantly records the position of the statepoint in the decoherence leaf.

Understanding the matrix elements as coordinates allows one to translate more directly between geometric visualization and density-matrix representations in different bases.  As shown in Fig.~\ref{fig_representations}(a), consider two representations of the Fig.~\ref{figBlochCircle}(c) state with 1/3 chance of being in the up/down/right pure states where the right statepoint is set to be at $\phi=0$ by convention.  Since we've located the statepoint geometrically, we see that if we choose the diameter for the up/down basis, the perpendicular of length 1/6 cuts the basis diameter in half, meaning its representation in this basis is $\begin{pmatrix} 
\frac{1}{2} & \frac{1}{6} \\
\frac{1}{6} & \frac{1}{2} \\
\end{pmatrix}$.  If we instead choose to find the matrix representation of this same state in the right-left basis, where the statepoint cuts the corresponding horizontal diameter directly in a $\frac{2}{3}$-to-$\frac{1}{3}$ ratio and the perpendicular projector length is 0, the matrix $\begin{pmatrix} 
\frac{2}{3} & 0 \\
0 & \frac{1}{3} \\
\end{pmatrix}$ represents the state.  While not all basis calculations are this straightforward, the geometric understanding provides a useful heuristic that can precede, accompany, or subsequently reinforce algebraic calculations of the representations.  Though algorithmic, the basis change alone in the standard algebraic formalism typically requires the construction of a unitary matrix to transform between bases and multiple matrix multiplications to produce the new matrix in contrast to the more intuitive and efficient geometric calculation shown here.

Note that, for any state and corresponding statepoint, the unique choice of basis where the off-diagonal elements are 0 is known as the diagonalized basis -- geometrically this basis is the unique diameter that passes through the statepoint.  The maximally mixed state of Fig.~\ref{figBlochCircle}(d), such as corresponds to an unpolarized photon state or one spin-1/2 particle in an entangled singlet state, has the statepoint at the center of the circle and is represented by $\begin{pmatrix} 
\frac{1}{2} & 0 \\
0 & \frac{1}{2} \\
\end{pmatrix}$ in any basis. 

While both pure and mixed states can be expressed in any basis as a density matrix with entries indicating the corresponding statepoint location relative to the basis diameter, there exists an alternative pure-state coordinate formalism for locating a statepoint on the boundary $\ket{\Psi}$ relative to the basis endpoint pure states $\ket{A}$ and $\ket{B}$.  The formula $p=\sqrt{1-a^2}=\sqrt{1-\left|\braket{A|\Psi}\right|^2}$ relates the basis-invariant distance $a\equiv\left|\braket{A|\Psi}\right|$, calculated from the states' inner product, between any two pure statepoints and the probability $p$ that $\ket{\Psi}$ will turn into $\ket{A}$ when a projective measurement is done in the $\ket{A}$, $\ket{B}$ basis.  To locate any pure statepoint relative to two basis statepoints one uses the column vector representation $\begin{pmatrix} 
a \\
b \\
\end{pmatrix}\equiv
\begin{pmatrix} 
\left|\braket{A|\Psi}\right| \\
\left|\braket{B|\Psi}\right| \\
\end{pmatrix}
$ where $a$ is the distance to one statepoint and $b$ to the other.  This locates the statepoint only up to a circular cross section of the Bloch sphere that bounds the decoherence leaf it is in relative to the chosen basis.  A unique specification of the state requires the specification of the angle $\phi$ of the state point about the measurement diameter.  This angle is encoded in the column vector representation as a relative phase between the two distances, $\begin{pmatrix} 
a \\
be^{i\phi} \\
\end{pmatrix}$,
 which is the most common way of representing pure qubit states. The column vector entries also specify the state as a basis-independent superposition (linear combination) of the basis-independent ket notation for the states $\ket{\Psi}=a\ket{A}+be^{i\phi}\ket{B}$.  These relations between pure statepoint distances, probabilities, and inner products applies to any dimensional statespace and are derived in Section~\ref{SecPure}.

\section{Beyond the Bloch Sphere:  Probability Simplices}
 \begin{figure}[h!]
\centering
\includegraphics[scale=.78]{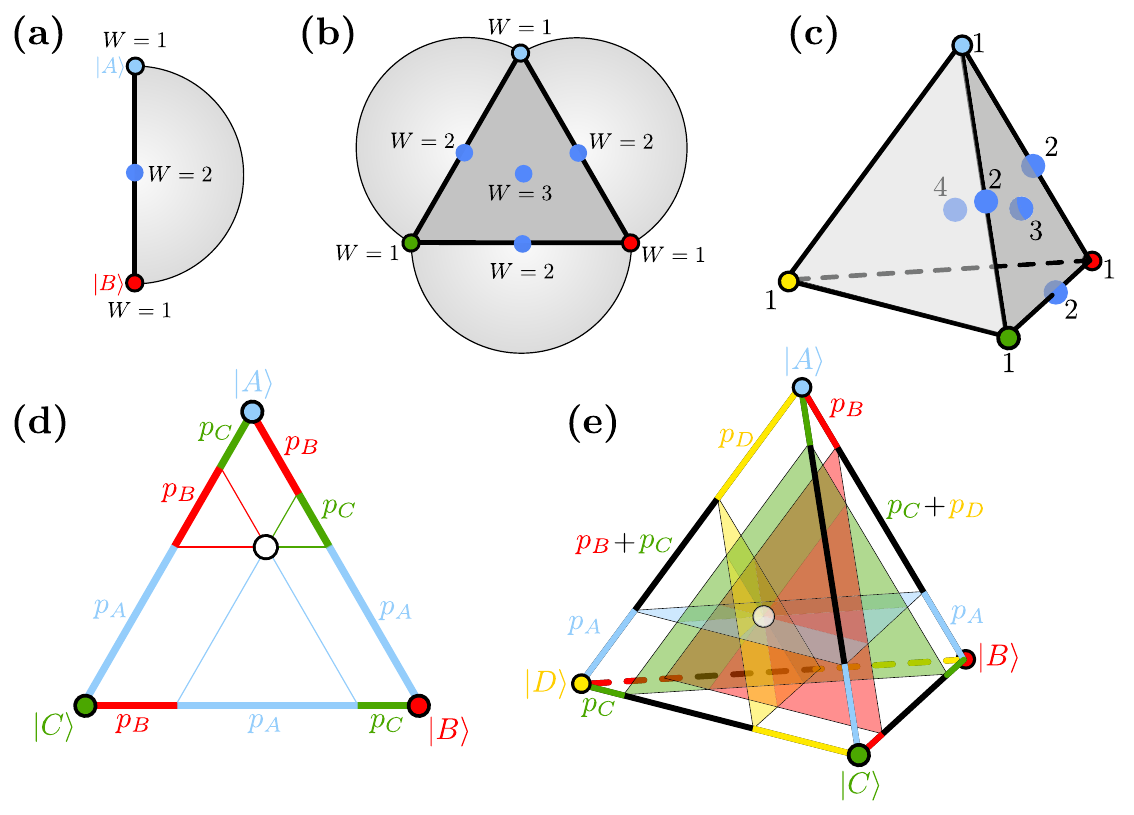}
\caption{(a) A regular, unit 1-simplex diameter of a partially-shown Bloch Sphere terminating at two zero-entropy pure states and bisected by the $d=2$ maximally mixed state with entropy $S=k_B\log{W}=k_B\log{2}$.  All qubit statepoints lie along one such diameter. (b) A regular, unit 2-simplex equilateral triangle with sides formed by qubit 1-simplices with the maximally mixed $W=3$ statepoint at its center.  The three Bloch Spheres corresponding to these 1-simplices are embeded in the 8-dimensional qutrit statespace where they intersect 2-simplices only at their edges and touch each other only at a single pure statepoint.  (c)  The regular, unit 3-simplex cross-section of the $d=4$ statespace with the $W=4$ maximally mixed statepoint at its center.  The tetrahedron's sides are distinct qutrit 2-simplices and its edges the qubit 1-simplices. (d) All qutrit statepoints lie in one of the 2-simplices; drawing lines through this statepoint parallel to the edges cuts the unit edges in lengths equal to the diagonalized density matrix measurement probabilities. An interactive version of this diagram is found \href{https://quantum.bard.edu/singleAnimation/qutritsimplex.html}{here}.  (e)  All $d=4$ statepoints lie in one of the 3-simplices; drawing planes through this statepoint parallel to the sides cuts the unit edges in lengths equal to the diagonalized density-matrix measurement probabilities and their sums. An interactive version of this diagram is found \href{https://quantum.bard.edu/singleAnimation/2qubitsimplex.html}{here}.}
\label{figSimplices}
\end{figure}
The qubit geometry contains within it several properties that are intuitive in the 3-dimensional Bloch sphere statespace, but also hold for the geometry of quantum states in any dimension.  To avoid confusion between the geometrical dimension of the real, Euclidean statespace and the quantum dimension of the number of basis states needed to specify the state, we use the index $n$ for the former ($n=3$ for the Bloch sphere representation of a qubit) and the index $d$ for the latter ($d=2$ states form a qubit basis).  A qutrit with $d=3$ is the next highest basis number, such as specifies the angular momentum of a spin-1 particle, and is represented in an $n=8$ dimensional statespace.  In general a ``qudit'' has a $d$-dimensional basis and is represented by a point in an $\left(d^2-1\right)=n$-dimensional statespace (see Appendix~\ref{App_ProbSimplex} for proof of this relation).  While this eight-dimensional Euclidean space in which each qutrit state corresponds to a statepoint is unvisualizable, we will see that multiple intuitive notions discussed in the qubit space are extendable to this higher-dimensional space.

To begin with, consider this geometric fact about the spherical qubit statespace: for each possible statepoint in the sphere there exists a unit diameter through that statepoint that passes through the maximally mixed statepoint at its center and terminates on the surface in two orthogonal pure statepoints.  This statement is a manifestation of a more general fact about qudit statepoint geometry:

\textit {For each possible statepoint in quantum dimension $d$ there exists a regular unit $d-1$ simplex that passes through the maximally mixed statepoint at the center of the $n=d^2-1$ statespace and terminates on the extremal boundary of the statespace at $d$ orthogonal pure statepoints.}

As shown in Fig.~\ref{figSimplices}, for the $d=2$ qubit space a ``regular unit $(2-1)=1$ simplex'' is a Bloch sphere unit diameter.  In the $d=3$ qutrit space a ``regular unit 2-simplex'' is an equilateral triangle with unit sides.  In the $d=4$ space a ``regular unit 3-simplex'' is a tetrahedron with unit sides.  Thus while we cannot visualize the full $d=3$ or $d=4$ statespaces, we can visualize a particular subspace of them for each possible statepoint.  In the $n=8$-dimensional qutrit space, each state will have a statepoint lying within one of these visualizable 2-simplex triangles.  The vertices of these simplices are the pure statepoints corresponding to a basis in which the density-matrix representation of the corresponding state is diagonalized, with the real, positive eigenvalues $p_1,p_2, . . . , p_d$ along the diagonal corresponding to the probabilities that the state will evolve to one of these $d$ states on a standard projection-value measurement in this basis.  In the case of the qubit, the placement of a statepoint on the 1-simplex unit diameter subspace cuts this diameter into two lengths that are equal to the $p_1, p_2$ measurement outcome probabilities.  More generally: 

\textit {A line/plane/hyperplane, passing through a statepoint and parallel to one side of the $d-1$ simplex it's located in, cuts the unit edges of the simplex to a length $p$ equal to the probability this state evolves on measurement to the pure state corresponding to the vertex opposite the chosen simplex side.}

Fig.~\ref{figSimplices}(d) shows how this statement works in the qutrit case.  Pick any statepoint in the triangle 2-simplex and draw lines through it parallel to the 2-simplex triangle sides.  These lines cut the unit edges in three pieces.  The length of a piece farthest away from the a vertex on that side is equal to the probability the chosen state would end up at that pure vertex statepoint on making a measurement in this basis.  As with the diameter being cut by a statepoint in the qubit case, the fact that it is the line segment farthest from the pure statepoint that represents the probability of ending up at that statepoint on measurement correlates with the intuitive fact that the closer one of the statepoints is to one of the pure state vertices, the more likely it will evolve to that state on measurement.  As shown in Fig.~\ref{figSimplices}(e), in the $d=4$ case it is now planes parallel to the sides of the tetrahedron that are drawn through the chosen statepoints, but the segments of the unit edges cut off by these planes are still equal to measurement probabilities for the opposite vertex pure state.  In general, simplex edges are cut into three pieces by this process; the two pieces at the end are equal in length to the probabilities for outcomes corresponding to the two vertices of that edge while the remaining piece in the middle is equal in length to the sum of the probabilities of all other outcomes.  As discussed further in Appendix~\ref{App_ProbSimplex}, these projections of statepoints on simplex edges using parallels in order to find probabilities are similar to reading off the coordinate parameters of a point using regular Cartesian axes, but with the coordinate axes intersecting at 60 degrees rather than 90 degrees.  The transformations there detailed between simplices and Cartesian coordinates are useful for proving relations between statepoints of qudits in any dimension.

The fact that each probability simplex is built up from lower-dimensional probability simplices is one manifestation of the fact that statespaces of higher dimensions contain the lower-dimsional statespaces within them.  Consider the n=8-dimensional statespace corresponding to all qutrit states.  As shown schematically in Fig.~\ref{figSimplices}(b), one subspace of this space is a 2-simplex triangle with three vertices corresponding to the pure states $\ket{A}, \ket{B}$, and $\ket{C}$.  Another subspace is all the qubit states found on the Bloch sphere that has states $\ket{A}$ and $\ket{B}$ as a basis.  The 2-simplex triangle and this Bloch sphere intersect at \textit{only} the line of statepoints directly joining the $\ket{A}$ and $\ket{B}$ statepoints.  In the simplex, this line is one edge of the triangle; in the Bloch sphere, this line is a diameter of the sphere. Two other Bloch spheres with bases $\ket{A}$, $\ket{C}$ and $\ket{B}$, $\ket{C}$ have diameters coincident with the other two triangle sides.  Pairs of each of these three Bloch spheres intersect each other only at the single pure statepoint they share.  Building to higher dimensions, the 3-simplex tetrahedrons in the $d=4$ statespace have four 2-simplexes as surfaces, which each exist in separate, but touching $d=3$ statespaces.  The unvisualizable 4-simplexes in the $d=5$ statespace will have five 3-simplexes joined to form its surface, and in general the $d-1$ simplexes in the $d$-dimensional statespaces will have $d$ number of $(d-2)$-simplexes joined to form their surface.

With this nesting sequence of statespaces in mind, a useful parameter with which to label statepoints is their entropic number of states $W$ linking the Boltzmann and von Neumann entropies $S=k_B\ln{W}=-k_B\rm{Tr}\left({\rho{\ln{\rho}}}\right)$.  All pure states, such as correspond to Bloch sphere surface points or simplex vertices, have zero entropy and $W=1$: they are perfectly described states with no possibility of more information being known about them.  The maximally mixed statepoint $\hat\rho_{M,2}$ at the center of a qubit and the midpoint of a simplex edge has $W=2$.  The maximally mixed qutrit state $\hat\rho_{M,3}$ at the center of the triangle or middle of a tetrahedron-face simplex has $W=3$, and in general the maximally mixed state $\hat\rho_{M,d}$ at the center of the $d$ statespace has $W=d$.

Since any statepoint can be found in a simplex with the maximally mixed statepoint corresponding to $\hat\rho_{M,d}$ at its center, one can use Euclidean geometry to calculate the distance between the statepoint and this center.  From the simplex probability geometry, the location of the statepoint within this simplex is determined by the measurement probabilities $p_i$, which are the entries in the diagonalized matrix representing the state in the basis corresponding to this simplex.  As shown in Appendix \ref{App_ProbSimplex}, the distance $r_d$ from this statepoint to the simplex center is given by $r_d^2=\frac{1}{2}\sum_{i=1}^{d} \left(p_i-\frac{1}{d}\right)^2$.  Somewhat more suggestively, the fact that the maximally mixed state in $d$ quantum dimensions $\hat\rho_{M,d}$ is represented by the diagonalized matrix $\rho_{M,d}$ with identical $\frac{1}{d}$ entries along the diagonal in any basis means this distance equation to any statepoint in the simplex represented by the the density matrix $\rho$ can be written $r_d^2=\frac{1}{2}\|\rho-\rho_{M,d}\|^2$.  Here the double vertical bars are the (Frobenius or Hilbert-Schmidt) norm of the resulting matrix -- the square root of the sum of the absolute value squares of each entry.  This result can be extended to find that the Euclidean distance $r_{ab}$ between any two statepoints in a simplex, corresponding to density matrices ${\rho}_a$ and $\rho_{b}$, is given by $r_{ab}^2=\frac{1}{2}\|\rho_a-\rho_{b}\|^2$, or, in terms of the measurement probabilities, $r_{ab}^2=\frac{1}{2}\sum_{i=1}^{d} \left({p_a}_i-{p_{b}}_i\right)^2$.  This quadratic relation reinforces the idea of the $p$'s as coordinate components, with the distance between the statepoints related to the sum of the square of the distances between each component, and leads to this general rule:  \textit{Each possible qudit basis corresponds to a unique regular unit $d-1$ simplex in the statespace with its $d$ vertices being statepoints corresponding to orthogonal basis states.  Any statepoint on this simplex is represented in that simplex basis as a diagonalized matrix whose diagonal element probabilities give its position within the simplex.}

\section{Qudit Decoherence Leaves}

The previous section outlined the geometry of the simplex bases and the locations of statepoints chosen within these simplices, but what about the general case of statepoints that do not lie within the chosen simplex?  As we present in this section, the entries of the density matrices of any state, written in the chosen basis, can be understood as Euclidean coordinates for the statepoint:  The diagonal elements of the matrix specify the coordinate location of the perpendicular projection of the statepoint onto the basis simplex, and the off-diagonal elements specify the length and direction of this perpendicular projector.

\begin{figure*}
	\centering
\includegraphics[scale=.72]{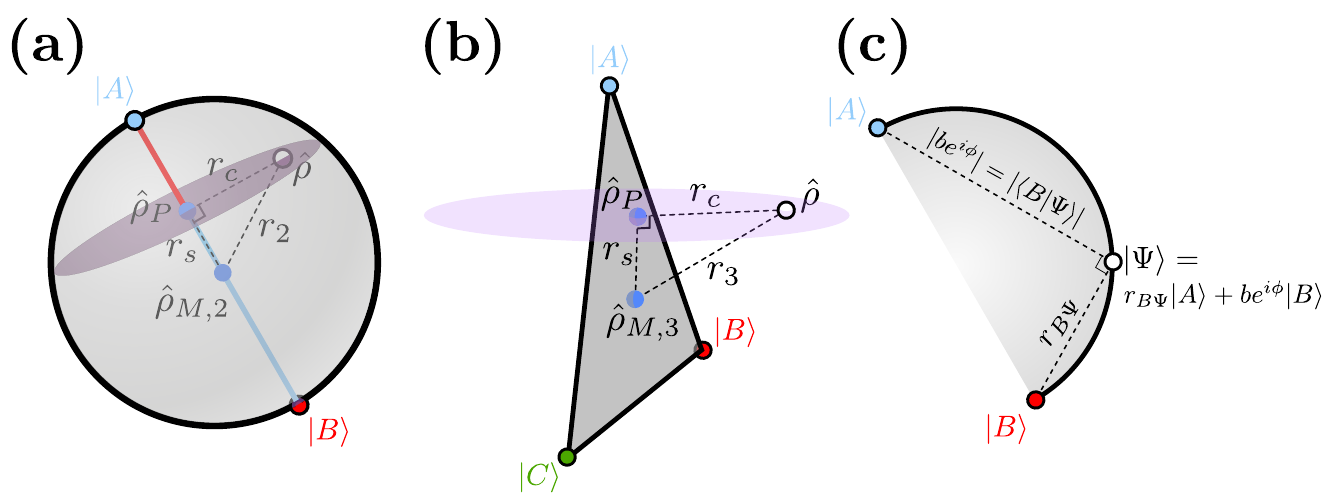}
\caption{For any state $\hat\rho$ and any choice of basis, the line $r_c$ in the decoherence leaf of the basis containing the $\hat\rho$ statepoint and its projection $\hat\rho_P$ statepoint on the basis simplex is perpendicular to the simplex.  Connecting these two statepoints to the maximally mixed statepoint $\hat\rho_{M,d}$ at the center of the simplex forms a right triangle as shown in the (a) $d=2$ qubit case where the leaf is 2-dimensional, and (b) $d=3$ qutrit case where the leaf is 6-dimensional.  (c) The length of line $r_{B\Psi}$ between any two pure states $\ket{B}$ and $\ket{\Psi}$ on the extremal boundary of a statespace is equal in length to a probability amplitude for the state $\ket{\Psi}$ expanded in a qubit basis containing $\ket{B}$.  Lines from the basis states to the statepoint form an alternate geometric parametrization of its location used in column- and row-vector representations of pure states.}
\label{fig_Projection}
\end{figure*}

To build intuition, we first recall the geometry of the Bloch sphere.  A choice of basis corresponds to a choice of diameter on the sphere (a 1-simplex).  Choosing any other statepoint in the sphere, we take the perpendicular projection of this point onto the chosen basis diameter.  As shown in Fig.~\ref{fig_Projection}(a) the location of this projection is a statepoint $\hat\rho_P$ on the unit diameter, which cuts it into two lengths.  These lengths give the diagonal element of the state's density matrix in that basis, reflective of the measurement probabilities in that basis.  The set of all statepoints in the statespace with this same projection statepoint $\hat\rho_P$ forms the decoherence leaf for that point on the diameter simplex.  The states corresponding to these points are all the states with identical diagonal density matrix elements in the chosen basis.  All the statepoints in a decoherence leaf form a circular cross-section of the Bloch sphere, perpendicular to the chosen basis diameter, and intersecting it at the projection point.  By specifying only the basis diameter and projection point $\hat\rho_P$, we have selected one decoherence leaf of that basis and narrowed down the possible statepoints to one within this leaf.  The particular statepoint within this circular cross-section can be specified in polar coordinates using the single complex number $r_ce^{i\phi}$ (Fig.~\ref{figMeasurement}(b)), which is the off-diagonal element of the density matrix representing the state.  As shown in Fig.~\ref{fig_Projection}(a) the radial distance $r_2$ from the maximally mixed statepoint $\hat\rho_{M,2}$ at the center of the Bloch sphere to any statepoint is the hypotenuse of a right triangle.  One leg of the triangle is the $r_s$ distance from $\hat\rho_{M,2}$ to the projection point $\hat\rho_P$ on the chosen basis diameter simplex; the other is the $r_c$ distance of the perpendicular projector from the statepoint to $\hat\rho_P$.

The virtue of understanding the geometry in the qubit case is that it extends to the general qudit case in a straightforward way.  As shown schematically in Fig.~\ref{fig_Projection}(b) for the qutrit case, the chosen basis is a visualizable 2-simplex triangle with the three basis states as vertices.  For any statepoint $\hat\rho$ located in the unvisualizable $n=8$-dimensional Euclidean statespace, there exists a visualizable line through this point that is perpendicular to the simplex.  This perpendicular intersects the simplex at a projection point on it.  The length of the projection line, from the chosen $\hat\rho$ statepoint off the simplex to the $\hat\rho_P$ projection point on it, is $r_c$; the length from the simplex center $\hat\rho_{M,3}$ to the projection point is $r_s$; and, identical to the qubit case, the distance from the center to the chosen statepoint is $r_3=\sqrt{(r_s^2+r_c^2)}$.  One can find the distances $r_s$ and $r_3$ using the formulas of the previous section for distances between any two points on the same simplex since in the former case they are the distances between two points on the chosen basis simplex and in the later case one can use a different simplex that goes through the chosen $\hat\rho$ statepoint, which must include the maximally mixed statepoint $\hat\rho_{M,3}$ at the center of all simplices for the chosen dimesion $d$.

As in the qubit case, the density matrices of the statepoint and its projection point for the chosen basis have identical diagonal values, corresponding to identical measurement probabilities in that basis.  For a given projection point on a basis simplex we can define its decoherence leaf as all the statepoints with these diagonal values.  While this leaf was a two-dimensional circle in the qubit case, for the qutrit case each decoherence leaf is six-dimensional, and, in general is $d(d-1)$-dimensional.  In the qubit case, the two-dimensional polar-coordinate location of a statepoint within the circular decoherence leaf was given by the magnitude and argument of the single complex number that is the independent off-diagonal element of that state's density matrix in the chosen basis.  In the qutrit case the unique location of a point in this six-dimensional space is given by three sets of polar coordinates consisting of three lengths, $r_{c1}$, $r_{c2}$, and $r_{c3}$, and three angles $\phi_1$, $\phi_2$, and $\phi_3$, which can be encoded in three complex numbers, $r_{c1}e^{i\phi_1}$, $r_{c2}e^{i\phi_2}$, and $r_{c3}e^{i\phi_3}$.  As shown in Appendix~\ref{App_Decoherence}, these three complex numbers are the three independent off-diagonal elements of the 3 $\times$ 3 density matrix representing the state in the chosen basis.  While the bounded, six-dimensional decoherence leaf in which these three off-diagonal elements specify a particular point is not visualizable, one can understand the idea that each one is an independent coordinate with independent magnitude which increases from $0$ as the statepoint moves farther away from the simplex.  Furthermore, the total distance $r_c$ of the statepoint to the simplex is given by the familiar Euclidean relation $r_c^2=r_{c1}^2+r_{c2}^2+r_{c3}^2$.

The results of the preceding section and paragraphs should make clear how to understand the entries in a state's density matrix $\rho$ as geometrical coordinates locating its unique statepoint in the quantum statespace:  The choice of basis for the density matrix corresponds to a unique simplex in the statespace.  The diagonal elements of the density matrix are coordinates locating the perpendicular projection of the statepoint onto the simplex.  The off-diagonal elements are coordinates locating the specific statepoint in the decoherence leaf of all statepoints with the same projection.  If $\rho_P$ is the density matrix corresponding to the projective point, it will have no off-diagonal entries (true for all statepoints in the simplex) and diagonal elements equal to $\rho$ (true for all statepoints in the same decoherence leaf).  The difference $\rho-\rho_P$ is then a matrix with $0$'s on the diagonal and the off-diagonal elements equal to those of $\rho$, e.g. for the qutrit case:
\begin{align*}
\rho-\rho_P =  \begin{pmatrix} 
p_1 && r_{c1}e^{-i\phi_1} && r_{c2}e^{-i\phi_2}\\
r_{c1}e^{i\phi_1} && p_2 &&	r_{c3}e^{-i\phi_3}\\
r_{c2}e^{i\phi_2} && r_{c3}e^{i\phi_3} &&	p_3\\
\end{pmatrix} 
-\begin{pmatrix} 
p_1 && 0 && 0\\
0 && p_2 && 0\\
0 && 0 &&	p_3\\
\end{pmatrix}
=\begin{pmatrix} 
0 && r_{c1}e^{-i\phi_1} && r_{c2}e^{-i\phi_2}\\
r_{c1}e^{i\phi_1} && 0 &&	r_{c3}e^{-i\phi_3}\\
r_{c2}e^{i\phi_2} && r_{c3}e^{i\phi_3} &&	0\\
\end{pmatrix}
\end{align*}
The information in the off-diagonal elements is redundantly encoded in the matrices, with the complex conjugate of each upper-right entry also recorded in the lower left of the matrix.  Since the Euclidean distance $r_c$ from the $\rho$ to $\rho_P$ statepoints is found by summing the square of the magnitude squared of one set of these two copies, it can be written compactly using the matrix norm notation:  $r_c^2=\frac{1}{2}||\rho-\rho_P||^2$, where the $1/2$ factor accounts for the two copies.  This metric for the distance between these two statepoints not on the same simplex is identical to the one of the previous section between the statepoints on the same simplex and adds to the motivation to adopt it as a general metric for the statespace.  This metric exists in the literature, though typically without the $1/2$ factor that allows direct correspondence between density-matrix entry magnitudes and Euclidian distances and that appears in the qubit case as the choice of radius $1/2$ for the Bloch sphere.  With the generalized notion of the leaf radius $r_c$ of a statepoint to its simplex projection, the Pythagorean relation that held in the qubit $r_s^2+r_p^2=r_d^2$ case can be shown to hold for any dimension $d$.

Since the distance $r_{ab}$ between two statepoints is a basis-independent quantity, the general metric can also be written (Appendix~\ref{App_Decoherence}) in a basis-independent form for any two states $\hat\rho_a$ and $\hat\rho_b$ or their representing matrices $\rho_a$ and $\rho_b$ in any basis using the trace operation:
\begin{equation}
	r_{ab}^2=\frac{1}{2}||\rho_a-\rho_b||^2=\frac{1}{2}\mathrm{Tr}\left(\rho_a^2\right)+\frac{1}{2}\mathrm{Tr}\left(\rho_b^2\right)-\mathrm{Tr}\left(\rho_a\rho_b\right)=\frac{1}{2}\mathrm{Tr}\left(\hat\rho_a^2\right)+\frac{1}{2}\mathrm{Tr}\left(\hat\rho_b^2\right)-\mathrm{Tr}\left(\hat\rho_a\hat\rho_b\right)
	\label{Eq_distance}
\end{equation}
From this definition, one can prove the triangle equality for the metric and show, consistent with the barycentric intuition for the qubit case, that the statepoint corresponding to the mixture of two states, $\hat\rho_c=p_a\hat\rho_a+p_b\hat\rho_b$, lies on the straight line connecting the statepoints for $\hat\rho_a$ and $\hat\rho_b$, nearer to the more probable state in the mixture, and cutting this line in a ratio of the mixture probabilities $p_a$ and $p_b$ (Appendix~\ref{App_Decoherence}).  This relation also underpins the convex nature of quantum statespaces for mixed states:  for any two statepoints in a statespace, all points on the line that joins them are also statepoints in the statespace.

While many Bloch sphere properties provide useful, visualizable analogies for the general higher-dimensional statepaces, its geometry can be a misleading analogy when comparing the surface of the statespace to the set of all pure statepoints, which we term the ``extremal boundary'' of the statespace.  An $\left(d^2-1\right)=n$-dimensional statespace has a $\left(d^2-2\right)$-dimensional surface.  Picking any line through a point in the statespace (for example a point and line in the 2-simplex triangle cross-section of the qutrit space) one can continue along this line through other statepoints up until a statepoint on the surface is reached (for example at the edge of the 2-simplex triangle).  Continuing the line further will reach points outside the statespace that do not correspond to any quantum state.  However, as can be seen in the 2-simplex triangle example, it is in general unlikely that the statepoint reached at the surface by traveling along a random line corresponds to a pure state (one of the three vertices of the triangle).  The set of pure statepoints is a $\left(2d-2\right)$-dimensional subspace of the surface states, distinguished by the fact that they are the set of statepoints farthest from a maximally mixed state at the center of the statespace and as such set its extremal boundary (e.g. the vertices of the 2-simplex triangle are the three points farthest from its center).  The $d=2$ Bloch sphere statespace is the {\it only} one where the surface and extremal boundaries are the same dimension, $d^2-2=2d-2$, both being the surface of the sphere.  In all other statespace dimensions the surface and boundary are distinct with the latter being a lower-dimensional subset of the former.

Algorithmically, one can check if any state $\hat\rho$ is a pure state on the boundary by seeing if its \textit {purity} $\rm{Tr}\left(\hat\rho^2\right)$ is equal to one -- the geometrical significance of this metric is discussed in Section \ref{statevectors}.  A statepoint on the boundary in a dimension $d$ will also be on the boundary for all higher dimensional statespaces; however, this is not the case for statepoints on the surface, since all interior statepoints in a $d$-dimensional statespace become surface statepoints in higher dimensions (e.g. statepoints in the interior of a $d=3$ two-simplex triangle become points on the surface of a $d=4$ three-simplex tetrahedron).  Algorithmically, one can deterimine if a state in the $d$-dimensional statespace is on its surface if its $d\times{d}$ representing matrix $\rho$ satisfies $\det{\left(\rho\right)}=0$, where the determinant is specifically the $d$-dimensional one.  Such a state can always be found in a lower-dimensional statespace with a matrix representation satisfying $\det{\left(\rho\right)}\ne0$ unless it is a pure state, which has a $0$ determinant in all statespaces and whose statepoint is always on the surface.

\section{Pure-State Coordinates and Metrics}
\label{SecPure}

In considering only statepoints on the pure-state boundary of the larger statespace, the general metric arrived at in the previous section for the distance between two statepoints takes on a simplified form.  Given a pure state $\ket{B}$, the corresponding density operator is constructed by taking the outer product of the ket and bra of the state, e.g. $\hat\rho_B=\ket{B}\bra{B}$. Since bras and kets commute within the trace operation, the distance between statepoints for $\hat\rho_B$ and $\hat\rho_\Psi$ is given by Eq.~\ref{Eq_distance}:
\begin{align*}
r_{B\Psi}^2&=\frac{1}{2}\mathrm{Tr}\left(\ket{B}\bra{B}\ket{B}\bra{B}\right)+\frac{1}{2}\mathrm{Tr}\left(\ket{\Psi}\bra{\Psi}\ket{\Psi}\bra{\Psi}\right)-\mathrm{Tr}\left(\ket{B}\bra{B}\ket{\Psi}\bra{\Psi}\right)\\
&=\frac{1}{2}\mathrm{Tr}\left(\left|\braket{B|B}\right|^2\right)+\frac{1}{2}\mathrm{Tr}\left(\left|\braket{\Psi|\Psi}\right|^2\right)-\mathrm{Tr}\left(\left|\braket{B|\Psi}\right|^2\right)=1-\left|\braket{B|\Psi}\right|^2,
\end{align*}
where we have used the normalization of the pure states.  The line between the two pure statepoints thus has a length $r_{B\Psi}=\sqrt{1-b^2}$, where $be^{i\phi}\equiv\braket{B|\Psi}$ is the corresponding probability amplitude entry of the column vector representing the state $\ket{\Psi}$ in a basis containing $\ket{B}$, and consequently the square of this length is always equal to the probability of \underline{not} measuring $\ket{\Psi}$ to be in state $\ket{B}$ in any measurement containing the latter as a basis vector. The column- and row-vector representations of pure states then also have a geometric significance in the qudit statespace, being an alternate way of locating the corresponding statepoint using coordinate lines drawn directly from the basis statepoints rather than coordinate lines locating the projection of the statepoint down to the basis simplex, which are used in density matrices.  Though simultaneously visualizing all lines directly from a pure state to a set of basis statepoints and their relative orientations in a general qudit statespace is not possible, one can always reduce the visualization of any single line to a qubit case embedded in this statespace as follows:

From the given pure states $\ket{B}$ and $\ket{\Psi}$, construct the state $\ket{A}\equiv\frac{1}{r_{B\Psi}}\ket{\Psi}-\frac{be^{i\phi}}{r_{B\Psi}}\ket{B}$.  It is straightforward to show that $\braket{A|A}=1$, $\braket{B|A}=0$ and $\ket{\Psi}=r_{B\Psi}\ket{A}+be^{i\phi}\ket{B}$, so that $\ket{A}$ and $\ket{B}$ form a qubit orthonormal basis embedded in the larger qudit space and $\ket{\Psi}$ is a qubit state in this basis space.  The Bloch sphere surface containing these three states is embedded in the qudit statespace and so the $r_{B\Psi}$ Euclidean distance between the boundary points $\ket{B}$ and $\ket{\Psi}$ is then the same as the qubit case (Fig.~\ref{fig_Projection}(c)).  Following Fig.~\ref{fig_representations}(c), the argument $\phi$ of the complex probability amplitude $\braket{B|\Psi}$ encodes the angle about the 1-simplex diameter joining the statepoints for $\ket{A}$ and $\ket{B}$ at which the lines of length $b$ and $r_{B\Psi}$ are joined to locate the statepoint of $\ket{\Psi}$. In general, any two statepoints are separated by a distance in the statespace ranging from 0 (when the states are the same) to 1 (when the states are orthogonal pure states).

The succession of embeddings of qubit statespaces (Bloch Spheres) within higher-$d$ qudit spaces while preserving the same metric can be extended to the continuum limit where pure states and mixed states are represented by continuous functions $\psi(x)$ and $\rho(x,x')$ rather than vectors and matrices.  The distance between pure states $\ket{\psi}$ and $\ket{\chi}$ continues to be $\sqrt{1-\left|\braket{\psi|\chi}\right|^2}$, and the general distance metric for states $\rho_a(x,x')$ and $\rho_b(x,x')$ becomes
\begin{equation}
	{r_{ab}^2=\frac{1}{2}\int{\rho_a}^2(x,x)dx+\frac{1}{2}\int{\rho_b}^2(x,x)dx-\int\int{\rho_a}(x,x')\rho_b}(x',x)dxdx'.
\end{equation}
It is important to emphasize that this statespace distance is unrelated to any real-space distance between spatial states.  For example the real-space distance between two position eigenstates corresponding to objects with definite spatial locations at $x_A$ and $x_B$, represented by Dirac-delta wavefunctions, is $\left|x_A-x_B\right|$, while the statespace distance is always 1 for $x_A\neq{x_B}$, since any two orthogonal states are always separated by a statespace distance of 1, which is the maximum distance between any two statespace points in any dimension.  

\section{Euclidean Statevectors and Angles}
\label{statevectors}

\begin{figure*}
	\centering
\includegraphics[scale=.21]{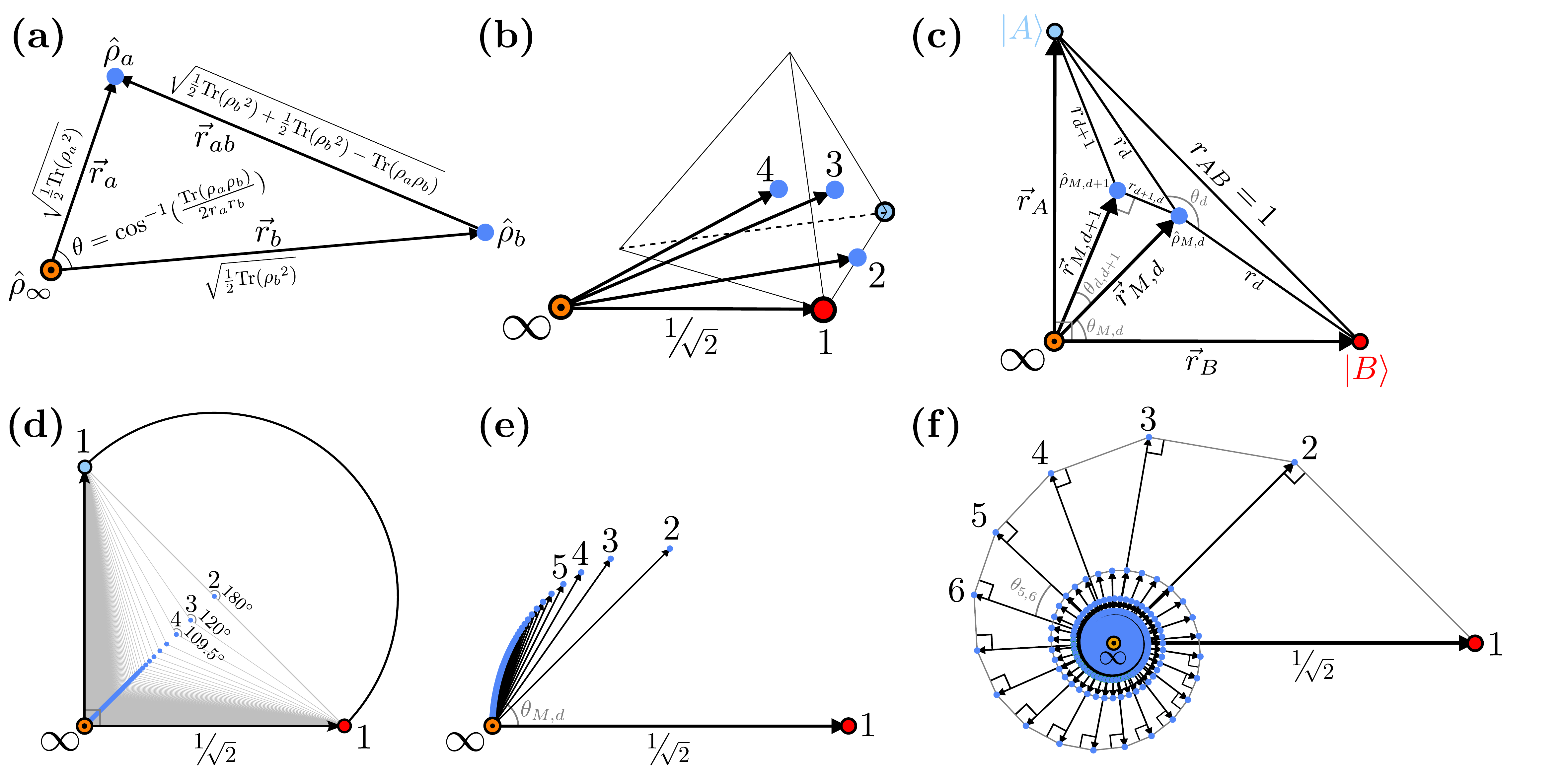}
\caption{(a) A triangle in the full statespace connecting statepoints of two general states $\hat\rho_a$, $\hat\rho_b$, and the maximally mixed state with infinite entropy $\hat\rho_\infty$.  (b) A schematic of the Euclidean statevectors connecting this $\hat\rho_\infty$ point with a pure state and maximally mixed states $\hat\rho_d$ for dimensions $d=2,\;3,\;\rm{and}\;4$ at the centers of the nested one-, two-, and three-dimensional probability simplexes.  Note that the $d=\infty$ origin point lies in dimensions outside this and all other finite-dimensional simplices, but is a finite distance $r_{M,d}$ from each of these statepoints.  (c) Statevectors from the $\hat\rho_\infty$ statepoint origin to two orthogonal pure statepoints, $\ket{A}$ and $\ket{B}$, and two successive maximally mixed statepoints, $\hat\rho_{M,d}$ and $\hat\rho_{M,d+1}$, that are mixtures of them. Triangles are formed by labeled lines connecting statepoints, but no two triangles lie in the same plane so distances and angles are shown projectively.  (d) Projectively shown triangles formed by the lines $r_d$ connecting $\hat\rho_{M,d}$ statepoints to two pure, orthogonal statepoints in their mixture.  All triangles share the same base: the diameter of the Bloch sphere connecting the pure statepoints (its half-perimeter shown schematically).  The distances from statepoints to the origin are drawn to scale, but the diameter-subtending angles $\theta_d$ are not, since each triangle is in a different plane. (e) Shows the correct $\vec{r}_{M,d}$ statevector lengths and the correct angle $\theta_{M,d}$ each makes with a pure statevector. (f) Accurately shows the $\vec{r}_{M,d}$ statevector lengths and the angles $\theta_{d,d+1}$ between \textit{successive} statevectors.  The right triangles formed by successive statevectors each lie in different plane as the previous one, but share a side.}
\label{figInfOrigin}
\end{figure*}

The fact that there is a one-to-one correspondence between qudit states and points in high-dimensional Euclidean spaces also means one can associate a real vector with every state by choosing an origin in these spaces.  Such a Euclidean statevector should not be confused with the more abstract column and row vectors with complex entries typically used to represent only pure states, as the magnitude and direction of the Euclidean statevector contains equivalent information to density matrix entries that can represent both pure and mixed states.\cite{kimura}  The simplest visualizable example of such a vector is the ``Bloch vector'' for the qubit space, discussed in multiple quantum textbooks, where one chooses the center of the Bloch sphere as the origin and associates the vector from this origin to the statepoint with the quantum state.  For a general qudit state of dimension $d$, the statepoint corresponding to the maxmimally mixed state, $\hat\rho_{M,d}$, is a natural choice of origin as it lies at the center of every probability simplex cross-section and is equidistant from all pure statepoints.  The real vector drawn from this statepoint to a general state $\hat\rho$ in its statespace is the $\left(d^2-1\right)$-dimensional ``generalized Bloch vector'' $\vec{r}_d$ with a length
\begin{equation}
	r_d^2=\frac{1}{2}\mathrm{Tr}\left(\rho^2\right)-\frac{1}{2d}.
\end{equation}
This equation and other quoted results in this section can be derived from the general distance metric of Eq.~\ref{Eq_distance} and basic triangle geometry, therefore making for good exercises for those beginning their understanding of mixed states and density matrices.  Full derivations are found in Appendix~\ref{App_Statevector}.
It follows that in the limit $d\rightarrow\infty$ this distance asymptotically approaches a fixed value $r=\sqrt{\left(\frac{1}{2}\mathrm{Tr}\left(\rho^2\right)\right)}$, which is the distance from the $\hat\rho$ statepoint to the statepoint for the maximally mixed state of infinite dimensionality $\hat\rho_{M,\infty}\equiv\hat\rho_{\infty}$, the state of infinite entropy.  As shown in Fig.~\ref{figInfOrigin}, this statepoint, lying at the center of the full statespace in the infinite-dimensional limit, serves as a convenient origin for states of any dimension.  Whether such an infinite-entropy state corresponds to anything physical may be left an open question, since geometrically the statepoint of any maximally mixed state of high, but finite dimension $D$ can be taken as an alternate origin instead with distance errors of order $\frac{1}{\sqrt{2D}}$ compared to the infinite limit.

Choosing the statepoint of infinite entropy as our origin and associating Euclidean state-vectors $\vec{r}$ from this origin of length $r=\sqrt{\left(\frac{1}{2}\mathrm{Tr}\left(\rho^2\right)\right)}$ to each statepoint simplifies the understanding of the terms in our general metric.  With $\vec{r}_a$ and $\vec{r}_b$ being the statevectors for states $\hat\rho_a$ and $\hat\rho_b$, Eq.~\ref{Eq_distance} becomes
\begin{equation}
	r_{ab}^2=r_{a}^2+r_{b}^2-\mathrm{Tr}\left(\rho_a\rho_b\right),
	\label{Eq_distance_2}
\end{equation}
from which it is apparent (see Fig.~\ref{figInfOrigin}(a)) that the metric corresponds to the law of cosines applied to the triangle connecting the statepoints of $\hat\rho_{\infty}$, $\hat\rho_a$, and $\hat\rho_b$ with vector sides $\vec{r}_a$, $\vec{r}_b$, and $\vec{r}_{ab}=\vec{r}_a-\vec{r}_b$, leading to $2\vec{r}_a\cdot\vec{r}_b=\mathrm{Tr}\left(\rho_a\rho_b\right)$.  This geometry allows the direct association of the final term with the dot product of the Euclidean statevectors and the determination of the angle $\theta_{ab}$ between them:
\begin{equation}
	\theta_{ab}=\cos^{-1} \left(\frac{1}{2r_{a}r_{b}}\mathrm{Tr}\left(\rho_a\rho_b\right)\right).
	\label{Eq_angle}
\end{equation}
From these triangles, the angle between any two lines connecting any three statepoints can then be determined using Euclidean geometry.

While the infinite dimensionality of the full statespace is a barrier to visualization, the use of lines connecting two statepoints and triangles connecting three statepoints are useful tools for building spatial intuition about the larger space.  A useful infinite subset of statepoints to consider in the full statespace is the sequence corresponding to maximally mixed states $\hat\rho_{M,d}$ that are mixtures of successively higher numbers of pure states and lie at the centers of the $\left(d^2-1\right)=n$-dimensional subspaces and $d-1$ dimensional simplices embedded within it (Fig.~\ref{figInfOrigin}b).  As shown in Fig.~\ref{figInfOrigin}(c), these statepoints are a distance $r_{M,d}=\sqrt{\frac12\,\operatorname{Tr}\!\left(\rho_{M,d}^{2}\right)}=\frac{1}{\sqrt{2d}}$ from the infinite-entropy origin that they asymptotically approach, with the $d=1$ pure statepoints a distance $\frac{1}{\sqrt{2}}$ from it bounding the statespace. This asymptotic approach to the $\hat\rho_\infty$ origin is shown in Fig.~\ref{figInfOrigin}(d), which also denotes the progression of the angle $\theta_d=\cos^{-1}\!\left(\frac{1}{1-d}\right)$ between lines connecting each statepoint $\hat\rho_{M,d}$ and two orthogonal pure states in the mixture.  $\theta_2=180^\circ$ is the familiar fact that these two lines form the straight diameter of a Bloch sphere with $\hat\rho_{M,2}$ at its center. $\theta_3=120^\circ$ is the angle between lines from the center of the 2-simplex equilateral triangle to two of its vertices, and $\theta_4=109.5^\circ$ is the angle between lines from the center of the 3-simplex tetrahedron to two of its vertices, etc.  It is remarkable that, in the $d\rightarrow\infty$ limit, $\theta_\infty=90^\circ$ is the angle between the Euclidean statevectors for the two pure states.  While the conventional understanding of two such states being ``orthogonal'' relies on an abstract notion of orthogonality based on an inner-product defined over the complex numbers, we here see that any two orthogonal states are located in the statespace by two real-space vectors that are at right-angles to each other, restoring the conventional meaning of the term as first encountered in Euclidean geometry.

Alternate methods of showing the progression of maximally mixed statepoints are shown in the last two panels of Fig.~\ref{figInfOrigin}.  Fig.~\ref{figInfOrigin}(e) shows the correct length of each of the statevectors and the correct angle $\theta_{M,d}=\cos^{-1}\!\left(\frac{1}{\sqrt d}\right)$ each makes with a pure statevector that is part of each mixture.  There is a different Euclidean orthogonality apparent here in the $d\rightarrow\infty$ limit with this angle approaching $90^\circ$ as the statevectors diminish to length zero.  Fig.~\ref{figInfOrigin}(f) shows the same set of vectors with the correct angles $\theta_{d+1,d}=\cos^{-1}\!\left(\sqrt{\frac{d}{d+1}}\right)$ between successive statevectors.  A third Euclidean orthogonality is revealed in the construction of right triangles connecting successive maximally mixed points $\hat\rho_{M,d}$ and $\hat\rho_{M,d+1}$ to the $\hat\rho_\infty$ origin.  We again emphasize that these triangles are not in same plane -- each successive statepoint is located in a new dimension outside those of the preceding statepoints.

It is worthwhile to step back from examining the geometry of these points outlining a path through the infinite-dimensional statespace to consider their epistemological significance:  The $W=1$ pure statepoint at the start of the path corresponds to a state of complete knowledge and zero entropy.  The $W=2$ statepoint is that of a perfect dilemma: the maximally mixed qubit state where any of two possible distinct outcomes are equally likely.  The $W=3$ statepoint is a perfect trilemma and so forth.  The successive points of increasing possibilities spiral down through the infinite dimensions of the bounded statespace towards the point of absolute ignorance at its center.

\section{Conclusion}

In this work, we have presented an overview of some of the simplest properties of geometric representations of quantum states that may be of use from the most introductory to the most advanced levels of quantum studies.  Central goals have been to show that geometric statespaces exist for any quantum state of any dimension; that despite the high-dimensionality of these statespaces, many aspects of them are intuitive and directly visualizable; that they can provide a framework to accompany the linear algebra formalism of quantum mechanics, in some cases allowing more efficient ways of doing calculations; that they give additional meaning to column-vector and density-matrix representations of states as being coordinates of statepoints; and that they allow one an additional tool to think about quantum states and their evolution in a basis-independent way.

Finally, we note that quantum mechanics is widely purported to be a theory that is both true and not understandable.\cite{feynman}  While such claims may be exaggerated, they highlight that there can be a distinction between verifying something is true and understanding it.  A commonly encountered example of this distinction is found when a physical or mathematical problem is translated into an algebraic formalism and a series of valid syntactical operations is performed by a human or machine to arrive at a true conclusion, but during which the steps performed lose much or all of their meaningful interpretation -- the algebra shows a result to be true, but the calculator has little intuition for {\it why} the result is true.  Contrast this experience to doing a similar problem geometrically, where a person is more likely to both literally and metaphorically ``see'' how and why each step is true as the proof progresses.  Such experiences indicate that geometric visualization of a problem is one potential avenue for improving understanding.  Our hope is that more widespread awareness of geometric quantum statespaces can play a similiar role in deepening the general understanding of quantum mechanics.

The authors have no conflicts to disclose.

\appendix

\section{Barycentric Geometry of Qubit States} \label{App_QubitBary}
\label{mixed}
\begin{figure}[h!]
\centering
\includegraphics[scale=.2]{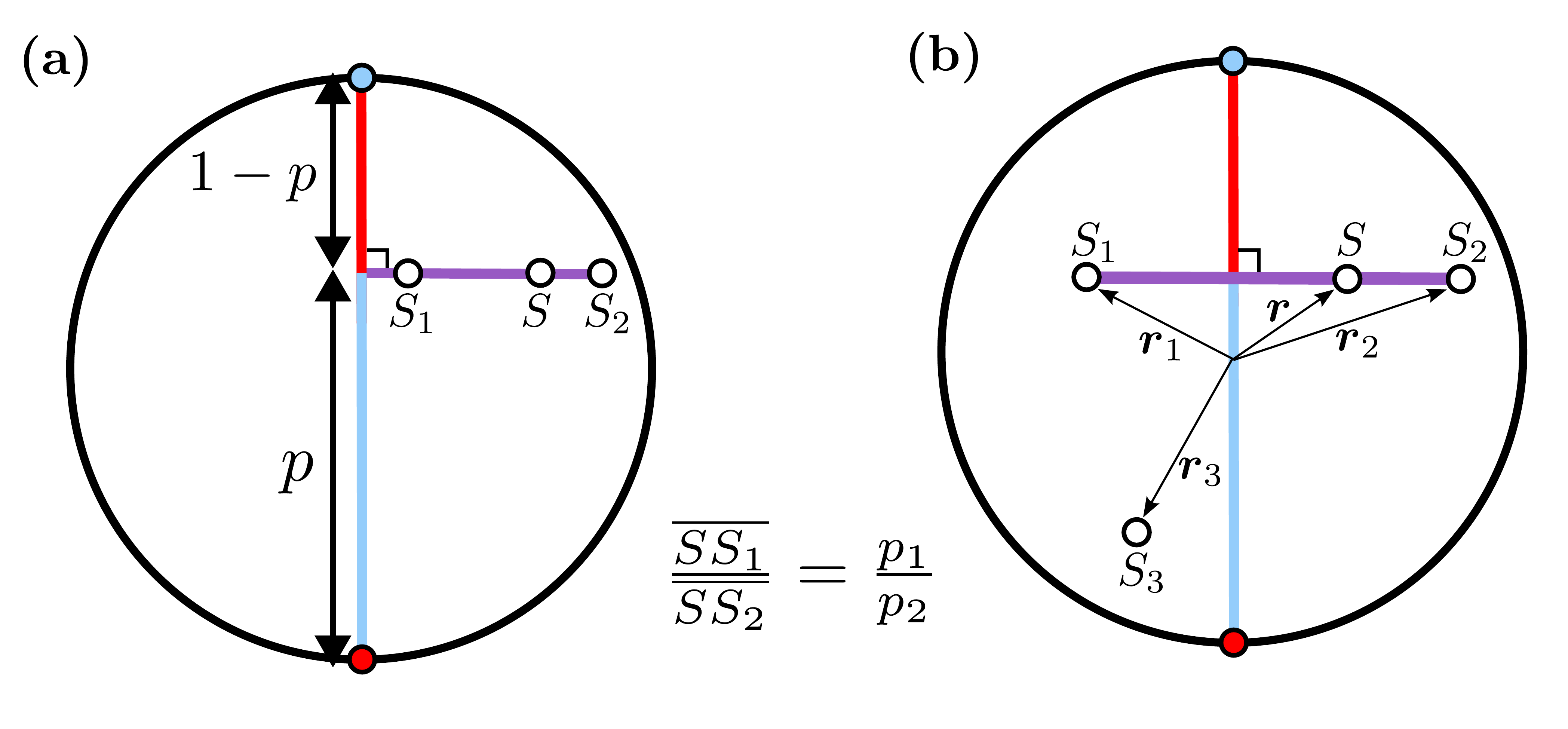}
\caption{Geometric picture of a linear combination of qubit mixed states using the epistemic probabilities $p_1$ and $p_2$ for the two possible cases.  With the states $\rho_1$ and $\rho_2$ plotted by the statepoints $S_1$ and $S_2$, the state $\rho=p_1\rho_1+p_2\rho_2$ is plotted by the statepoint $S$ that cuts the line segment joining $S_1$ and $S_2$ in the ratio of the epistemic probabilities.  As shown in (b) the location of statepoints can be specified by vectors drawn from the center of the Bloch Circle that satisfy the geometric relation $\boldsymbol{r}=p_1\boldsymbol{r}_1+p_2\boldsymbol{r}_2$, which has an identical algebraic structure to the corresponding density-matrix sum.}.
\label{figLinComb}
\end{figure}
While the following Appendices prove relations for the general qudit case, it is useful to first understand the barycentric geometry of quantum mixed states using the fully visualizable geometry of qubit states. To begin with, consider any two qubit states $\hat\rho_1$ and $\hat\rho_2$ and corresponding statepoints $S_1$ and $S_2$ in the Bloch sphere. We wish to find the statepoint $S$ corresponding to the linear combination $\hat\rho = p_1\hat\rho_1 + p_2\hat\rho_2$, where the epistemic probabilities of each state $p_1$ and $p_2$ are normalized: $p_1 +p_2 = 1$. Without loss of generality we choose a basis diameter that is perpendicular to the segment $\overline{S_1S_2}$ drawn between $S_1$ and $S_2$, so that $\hat\rho_1$ and $\hat\rho_2$ are in the same decoherence leaf of the basis, and choose our $\phi=0$ reference angle in the leaf to be in the direction of $\overline{S_1S_2}$.  The first of these choices ensures that the matrix representations $\rho_1$ and $\rho_2$ of the states in this basis have the same diagonal entries and measurement probabilities; the second that their off-diagonal entries are real, $\rho_1 = \begin{pmatrix}
p & c_1 \\
c_1 & 1-p
\end{pmatrix}
$ and $\rho_2 = \begin{pmatrix}
p & c_2 \\
c_2 & 1-p
\end{pmatrix},$
with $c_2>c_1$.  The two possible geometries for this set-up are shown in Fig. \ref{figLinComb} where (a) shows the $c_1>0$ case and (b) the $c_1<0$ case.  The off-diagonal elements of $\rho_1$ and $\rho_2$ correspond to the distances from the statepoints to this diameter.\cite{chang} From our choice of basis and the sign conventions of these off-diagonal elements, the length of $\overline{S_1S_2}$ is $c_2 - c_1$. Using the normalization, the representing matrix $\rho$ of the mixutre $\hat\rho$ in this basis is
$$\rho = p_1\rho_1 + p_2\rho_2 = 
p_1 \begin{pmatrix}
p & c_1 \\
c_1 & 1-p
\end{pmatrix}
+ p_2\begin{pmatrix}
p & c_2 \\
c_2 & 1-p
\end{pmatrix}
= \begin{pmatrix}
 p & p_1c_1 + p_2c_2 \\
p_1c_1 + p_2c_2 & 1-p
\end{pmatrix}.$$
Since this resulting matrix also shares the same diagonal elements, and has a real off-diagonal element, its statepoint $S$ must lie along the $\overline{S_1S_2}$ segment as well, with its placement on this line determined by its off-diagonal element ${p_1c_1 + p_2c_2}$. Geometrically the distance from $S_1$ to this statepoint $S$ is 
\begin{align*}
\overline{S_1 S} &= p_1c_1 + p_2c_2 - c_1 \\
&= (1 - p_2)c_1 + p_2c_2 - c_1 \\
&= p_2(c_2 - c_1) \\
&= p_2 \overline{S_1 S_2}.
\end{align*}
Similarly, $\overline{S S_2}= p_1 \overline{S_1 S_2}$, so that $\frac{\overline{S_1 S}}{\overline{S S_2}}=\frac{p_2}{p_1}$. Thus the statepoint $S$ has cut the line between the two constituent points in proportion to the epistemic probability weights.  The geometric location of the resulting statepoint can also be found algebraically in terms of their displacement vectors from the center of the circle (see Fig. \ref{figLinComb}(b)) or any other choice of origin.  With $\Vec{r_1}$ and $\Vec{r_2}$ being the vectors for the $S_1$ and $S_2$ statepoints respectively, the displacement from $S_1$ to $S_2$ is $\Vec{r_2}-\Vec{r_1}$.  Using the geometrical result and the normalization condition, the statepoint $S$ is a fraction $\frac{p_2}{p_1+p_2}=p_2$ along this displacement starting from the $S_1$ statepoint at $\Vec{r_1}$, and so is at the coordinate $\Vec{r}=\Vec{r_1}+p_2\left(\Vec{r_2}-\Vec{r_1}\right)=p_1\Vec{r_1}+p_2\Vec{r_2}$.  This linear combination of the statepoint coordinates has the same algebraic structure as the corresponding density-matrix linear combination $\rho = p_1\rho_1 + p_2\rho_2$.  Moreover, when this result is written $\Vec{r}=\frac{p_1\Vec{r_1}+p_2\Vec{r_2}}{p_1+p_2}$ the coordinate algorithm mimics that of finding the center of mass of two bodies located at $\Vec{r_1}$ and $\Vec{r_2}$ with the usual weighting of these positions by masses $m_1$ and $m_2$ replaced by the probability weights $p_1$ and $p_2$.  In this formulation, the coordinate $\Vec{r}$ is unchanged if $p_1$ and $p_2$ remain in the same ratio even if they are not normalized.

This formal analogy between the linear combination of density matrices and the ``center of mass'' of their corresponding statepoints can be extended to a linear combination of any number of terms by recursively applying the above result for combining two density matrices.  To see how this recursion works, consider the density matrix $\rho_0 = p_1\rho_1 + p_2\rho_2 + p_3\rho_3$, where the probability weights $p_1$ and $p_2$ are in the same ratio as above, but are no longer normalized since now the normalization condition is $p_1+p_2+p_3=1$.  Factoring $\rho_0 = (p_1+p_2)\left[\frac{p_1}{p_1+p_2}\rho_1 + \frac{p_2}{p_1+p_2}\rho_2\right] + p_3\rho_3$, the term in brackets is the combination of $\rho_1$ and $\rho_2$ having normalized coefficients $\frac{p_1}{p_1+p_2}$ and $\frac{p_2}{p_1+p_2}$ in the ratio $\frac{p_2}{p_1}$ and so is the same $\rho$ state with statepoint $S$ located at $\vec{r}$ found above when combining only the two density matrices. Thus $\rho_0 = (p_1 + p_2)\rho + p_3\rho_3$.  With $\rho_0$ now written as the combination of two density matrices with normalized probabilities $(p_1 + p_2)$ and $p_3$, we can again use the two-density-matrix geometric result to find the statepoint location for $\rho_0$ at coordinate $\Vec{r_0}$. Since $\rho$ corresponds to $\Vec{r}$, and denoting the coordinate corresponding to $\rho_3$ as $\Vec{r_3}$, the linear combination for $\rho_0$ indicates that $\Vec{r_0}=(p_1+p_2)\Vec{r}+p_3\Vec{r_3}=(p_1+p_2)\left[\frac{p_1\Vec{r_1}+p_2\Vec{r_2}}{p_1+p_2}\right]+p_3\Vec{r_3}=p_1\Vec{r_1}+p_2\Vec{r_2}+p_3\Vec{r_3}.$  Again this result can be expressed in a center-of-mass form as  $\Vec{r_0}=\frac{p_1\Vec{r_1}+p_2\Vec{r_2}+p_3\Vec{r_3}}{p_1+p_2+p_3}$, where the denominator is unity for a set of normalized probabilities.  With successive applications of the same algorithm, the coordinate location of a mixed state can be extended to a general result for any number of statepoints in the composition: for a state $\rho=\sum_{i} p_i\rho_i$ composed of states corresponding to statepoints $\Vec{r_i}$ on a Bloch sphere with normalized epistemic probabilities $p_i$, the statepoint of $\rho$ is located at $\Vec{r}=\sum_{i} p_i\Vec{r_i}$.

\section{Probability Simplex Geometry}
\label{App_ProbSimplex}

Any quantum-state dimension-$d$ density operator $\hat{\rho}$, or any density matrix representation of it $\rho$, is positive semi-definite with trace 1, meaning it is Hermitian with positive real eigenvalues $p_1,p_2,p_3,\ldots{p_d}$ that sum to 1 and has a corresponding set of orthogonal pure eigenstates with density operators $\hat{\rho}_1,\hat{\rho}_2,\hat{\rho}_3,\ldots\hat{\rho}_d$.  The $\hat{\rho}$ density operator can then be written $\hat{\rho}=\sum_{i=1}^{d}{p_i\hat{\rho}_i}$ and its representation in this eigenbasis of pure states (the ``measurement basis'' corresponding to this state) is a diagonalized matrix with the $p_i$ eigenvalues along its diagonal.
\begin{align}
    \hat{\rho}\rightarrow\rho=\begin{pmatrix}
        p_1 & 0 & 0 & 0 & \cdots & 0\\
        0 & p_2 & 0 & 0 & \cdots & 0\\
        0 & 0 & p_3 & 0 & \cdots & 0\\
        0 & 0 & 0 & p_4 & \cdots & 0\\
        \vdots & \vdots & \vdots & \vdots & \ddots & \vdots\\
        0 & 0 & 0 & 0 & \cdots & p_{d}
    \end{pmatrix}\label{diagonalized_rho}
\end{align}
The eigenvalues are the probabilities of the state ending up in each of the $\hat{\rho}_i$ eigenstates on projective measurement with the trace condition corresponding to the probability normalization $\sum_{i=1}^{d}{p_i}=1$.

\begin{figure*}
	\centering
\includegraphics[scale=.46]{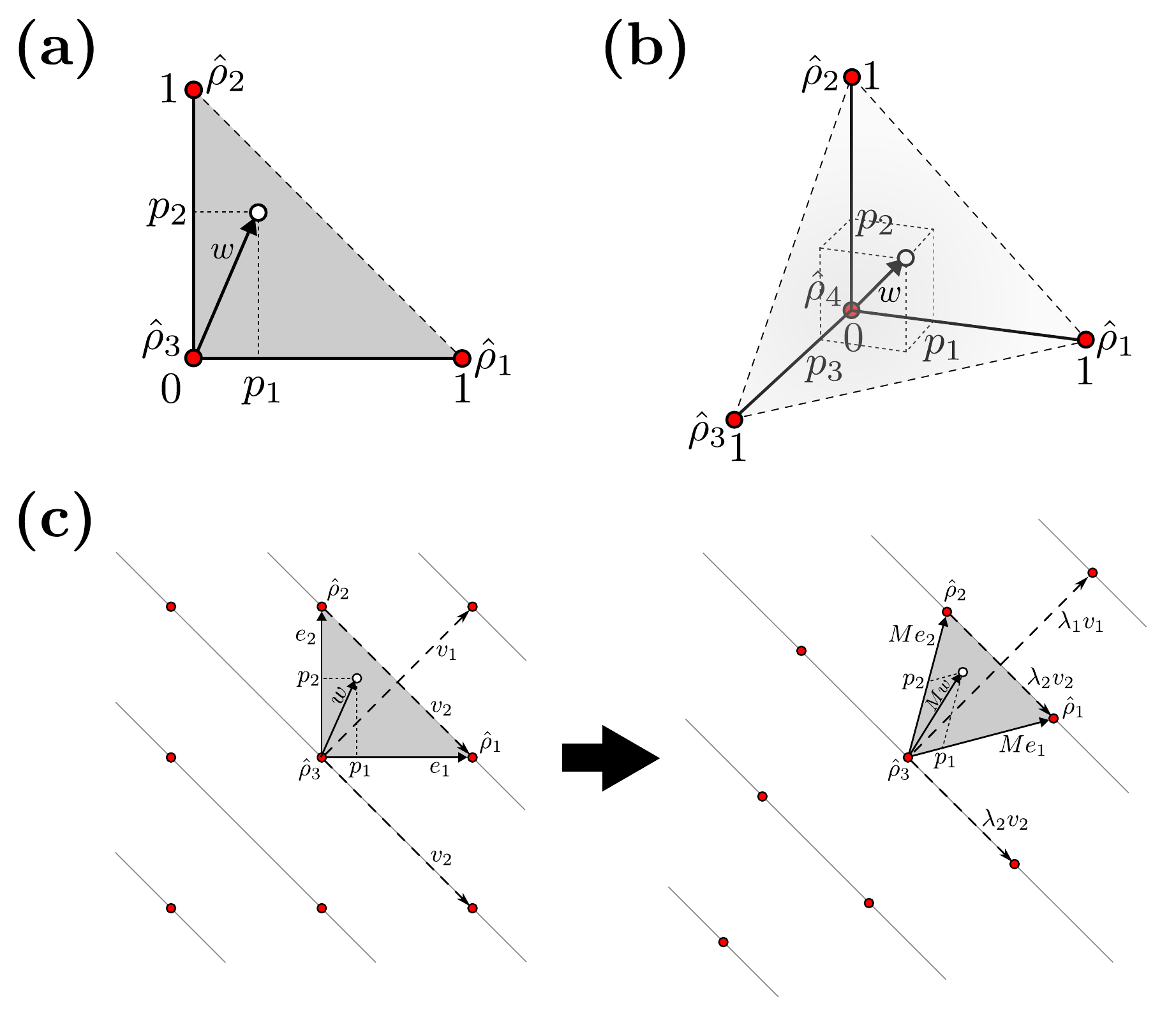}
\caption{(a) The irregular, Cartesian $d=3$ qutrit probability simplex where two of the three probabilities are coordinates of the statepoint.  (b) The irregular, Cartesian $d=4$ probability simplex where three of the four probabilities are coordinates of the statepoint.  (c)  The transformation of irregular simplices into regular ones uses a linear operation $M$ that turns a Cartesian grid into a triangular grid.  The transformation leaves the orientation of diagonal lines/planes/hyperplanes through next-nearest neighbors unchanged while increasing their spacing.  The unit vectors $e_i$ rotate towards the $v_1$ eigenvector that is perpendicular to the lines/planes/hyperplanes until the next-nearest neighbors become nearest neighbors.  The other eigenvectors of the transformation (e.g. $v_2$) lie in the diagonal lines/planes/hyperplanes.}
\label{figSimplexTrans}
\end{figure*}

Since every state can be written in this way in at least one measurement basis of orthogonal pure states, we chose one such basis and examine the subset of all states that can be written $\hat{\rho}=\sum_{i=1}^{d}{p_i\hat{\rho}_i}$ in that basis for all possible choices of the probabilities, i.e. all states represented by diagonalized density matrices in the chosen basis.  This subset of states is parameterized by the independent probabilities $p_1, p_2, p_3,\cdots{p_{d-1}}$, with the final probability $p_d$ then determined by the normalization condition.  As shown in Fig.~\ref{figSimplexTrans} the values of the ${d-1}$ independent probabilities can be plotted as the component coordinates of a point in a real $\left({d-1}\right)$-dimensional coordinate system. The probabilities must satisfy $0\leq\sum_{i=1}^{d-1}{p_i}\leq{1}$ with the 0 equality holding for the $\hat\rho=\hat\rho_d$ basis state, corresponding to the point at the origin.  Similarly each of the $d-1$ points a distance 1 from the origin along each axis (corresponding to orthonormal unit vectors $e_1,e_2,e_3,\cdots{e_{d-1}}$) are the statepoints for the other $\hat{\rho}_1,\hat{\rho}_2,\hat{\rho}_3,\ldots\hat{\rho}_{d-1}$ measurement basis states.

The upper-bound sum of one on the parametrizing probabilities $\sum_{i=1}^{d-1}{p_i}\leq1$ defines a linear equation for a geometric line/plane/hyperplane passing through the $\hat{\rho}_1,\hat{\rho}_2,\hat{\rho}_3,\ldots\hat{\rho}_{d-1}$ basis statepoints, such as is shown in Fig.~\ref{figSimplexTrans}(a) and (b) for the $d=3$ and $d=4$ case respectively.  Since each probability is positive or zero, this means that the diagonalized states in this basis are uniquely specified by a point in the volume bounded by the positive coordinate axes and this line/plane/hyperplane, which is an irregular $d-1$ simplex with vertices that are statepoints corresponding to the basis states.

The $p_1, p_2, p_3,\cdots{p_{d-1}}$ probabilities specifying each state can be read off in the usual manner by drawing lines parallel to the orthogonal Cartesian axes through the corresponding statepoint.  However, using Cartesian axes as edges of the irregular simplex creates asymmetries in the distances between points that are not reflective of the quantum state symmetries.  For example, though the basis state indexing is arbitrary, the distance of $\sqrt{2}$ between each of the $\hat{\rho}_1,\hat{\rho}_2,\hat{\rho}_3,\ldots\hat{\rho}_{d-1}$ statepoints is different from the unit distance between any of these and the $\hat{\rho}_d$ statepoint.  To remedy this assymmetry we perform a linear transformation on the irregular simplices to turn them into regular simplices with unit edges.

Fig.~\ref{figSimplexTrans}(c) shows the desired transformation in the $d=3$ case, where the square, Cartesian coordinate grid defined by the $e_i$ unit vectors is transformed into a triangular one by rotating these vectors towards a diagonal line along the direction $v_1\equiv{e_1+e_2}$.  The unit vectors maintain their unit length during the transformation, rotating until the statepoints at their ends are all separated by a unit distance so that the vector $v_2\equiv{e_1-e_2}$ shrinks in length from $\sqrt{2}$ to $1$.  A key property of this transformation in any dimension is that the lines/planes/hyperplanes, parallel to the one through the $\hat{\rho}_1,\hat{\rho}_2,\hat{\rho}_3,\ldots\hat{\rho}_{d-1}$ statepoints bounding the simplex, do not rotate during the transfomation, but do have their length scaled by this same factor. Thus in the general case the vectors $v_i=e_1-e_i$, for $i$ from $2$ to ${d-1}$ are eigenvectors of the transformation with degenerate eigenvalues $\lambda_2=\lambda_3=\lambda_4\cdots\lambda_{d-1}=\frac{1}{\sqrt{2}}$.  The diagonal vector, defined generally as $v_1\equiv\sum_{i=1}^{d-1}e_i$, is the perpendicular displacement between the parallel lines/planes/hyperplanes and is also a non-rotating eigenvector of the transformation.  Written in column vector form in the $e_i$ basis, these vectors form a linearly independent set in the $\left(d-1\right)$-dimensional space:
\begin{equation}
	v_1=\begin{pmatrix}
		1\\1\\1\\1\\{\vdots}\\1
		\end{pmatrix}\;\;
	v_2=\begin{pmatrix}
		1\\-1\\0\\0\\{\vdots}\\0
		\end{pmatrix}\;\;
	v_3=\begin{pmatrix}
		1\\0\\-1\\0\\{\vdots}\\0
		\end{pmatrix}\;\;
	v_4=\begin{pmatrix}
		1\\0\\0\\-1\\{\vdots}\\0
		\end{pmatrix}\;\;
	\cdots\;\;
	v_{d-1}=\begin{pmatrix}
		1\\0\\0\\0\\{\vdots}\\-1
		\end{pmatrix}\;\;
\label{Eq_Eigenvectors}
\end{equation}

The eigenvalue $\lambda_1$ corresponding to the $v_1$ eigenvector can be found by noting that this vector is the displacement from the origin to the $\left(d-1\right)^{th}$ nearest neighbor of the square/cube/hypercube lattice points and so has a length equal to $d-1$ multiples of the spacing between adjacent parallel lines/planes/hyperplanes.  Its length is $\sqrt{d-1}$, so the spacing between these parallels is $\frac{\sqrt{d-1}}{d-1}=\frac{1}{\sqrt{d-1}}$ before the transformation.  After the transformation the spacing is equal to the height of a regular $d-1$ simplex with unit sides, $\sqrt{\frac{d}{2(d-1)}}$, meaning the $v_1$ eigenvector is stretched by $\lambda_1=\frac{\sqrt{\frac{d}{2(d-1)}}}{\frac{1}{\sqrt{d-1}}}=\sqrt{\frac{d}{2}}$.

With all the eigenvalues and eigenvectors determined, we can find the the square-to-triangular lattice transformation matrix $M$ on the $e_i$ unit vectors by using Eq.~\ref{Eq_Eigenvectors} to express them in terms of the $v_i$ eigenvectors:
\begin{align*}
    e_1&=\frac{1}{d-1}\left(v_1+v_2+v_3\cdots+v_{d-1}\right)\\
    e_2&=\frac{1}{d-1}\left(v_1+v_3+v_4\cdots+v_{d-1}\right)-\frac{d-2}{d-1}v_2\\
    e_3&=\frac{1}{d-1}\left(v_1+v_2+v_4\cdots+v_{d-1}\right)-\frac{d-2}{d-1}v_3\\
    &\vdots\\
    e_{d-1}&=\frac{1}{d-1}\left(v_1+v_2+v_3\cdots+v_{d-1}\right)-\frac{d-2}{d-1}v_{d-1}\\
\end{align*}
and then using $Mv_i=\lambda_i{v_i}$ on these expansions, e.g.,
\begin{align*}
    Me_1&=\frac{1}{d-1}\left(\sqrt{\frac{d}{2}}v_1+{\frac{1}{\sqrt{2}}}v_2+\frac{1}{\sqrt{2}}v_3\cdots+\frac{1}{\sqrt{2}}v_{d-1}\right)=\frac{1}{\sqrt{2}\left(d-1\right)}\begin{pmatrix}
		\sqrt{d}+d-2\\\sqrt{d}-1\\\sqrt{d}-1\\\sqrt{d}-1\\{\vdots}\\\sqrt{d}-1
		\end{pmatrix}\;\;\\
    Me_2&=\frac{1}{d-1}\left(\sqrt{\frac{d}{2}}v_1+\frac{1}{\sqrt{2}}v_3+\frac{1}{\sqrt{2}}v_4\cdots+\frac{1}{\sqrt{2}}v_{d-1}\right)-\frac{d-2}{\sqrt{2}\left(d-1\right)}v_2=\frac{1}{\sqrt{2}\left(d-1\right)}\begin{pmatrix}
		\sqrt{d}-1\\\sqrt{d}+d-2\\\sqrt{d}-1\\\sqrt{d}-1\\{\vdots}\\\sqrt{d}-1
		\end{pmatrix}\;\;\\
\end{align*}
to find the columns of the transformation matrix
\begin{align}
    M=\frac{1}{\sqrt{2}\left(d-1\right)}\begin{pmatrix}
        \sqrt{d}+d-2 & \sqrt{d}-1 & \sqrt{d}-1 & \cdots & \sqrt{d}-1\\
        \sqrt{d}-1 & \sqrt{d}+d-2 & \sqrt{d}-1 & \cdots & \sqrt{d}-1\\
        \sqrt{d}-1 & \sqrt{d}-1 & \sqrt{d}+d-2 & \cdots & \sqrt{d}-1\\
        \vdots & \vdots & \vdots & \ddots & \vdots\\
        \sqrt{d}-1 & \sqrt{d}-1 & \sqrt{d}-1 & \cdots & \sqrt{d}+d-2
    \end{pmatrix}\label{Eq_Transformation_Matrix}.
\end{align}

With this matrix we can now transform any $\left(d-1\right)$-dimensional vector from the origin to a statepoint in the Cartesian coordinates $w=\begin{pmatrix}
		p_1\\p_2\\p_3\\{\vdots}\\p_{d-1}
		\end{pmatrix}$ to a new vector $Mw$ that points to the statepoint in the regular unit simplex (see Fig.~\ref{figSimplexTrans}(c)).  With the lengths of the $e_i$ unit vectors unchanged in the transformation as the angle between them goes from $90^\circ$ to $60^\circ$, the $p_i$ ratio by which each of them is cut by drawing a line/plane/hyperplane through the statepoint and parallel to the line/plane/hyperplane through the $d-1$ $\rho_k$ vertices with $k\neq{i}$ remains unchanged.
		
		While these coordinates do not change, the square of the length of any vector $w$ from the origin does.  In the Cartesian coordinates this length is given by the Euclidean inner product $w^Tw$.  After transformation the length of this vector in the unit simplex is $\left(w^TM^T\right)Mw=w^T\left(M^TM\right)w$.  From Eq.~\ref{Eq_Transformation_Matrix} the metric used to transform the square lengths between points in the Cartesian coordinates of the irregular simplex to the triangular coordinates of the regular simplex is
\begin{align}
    M^TM=\frac{1}{2}\begin{pmatrix}
        2 & 1 & 1 & \cdots & 1\\
        1 & 2 & 1 & \cdots & 1\\
        1 & 1 & 2 & \cdots & 1\\
        \vdots & \vdots & \vdots & \ddots & \vdots\\
        1 & 1 & 1 & \cdots & 2
    \end{pmatrix}.\label{Eq_Metric}
\end{align}

In the Cartesian coordinates, the distance between any two statepoints in the simplex for states $\hat\rho_b$ and $\hat\rho_c$ with respective probability coordinates $p_{b_i}$ and $p_{c_i}$ is the length of the vector
\begin{equation}
	\begin{pmatrix}
		\Delta{p_1}\\\Delta{p_2}\\\Delta{p_3}\\{\vdots}\\\Delta{p_{d-1}}
		\end{pmatrix}\equiv\begin{pmatrix}
		p_{b_1}-p_{c_1}\\p_{b_2}-p_{c_2}\\p_{b_3}-p_{c_3}\\{\vdots}\\p_{b_{d-1}}-p_{c_{d-1}}
		\end{pmatrix}.
\end{equation}
From the $w^T\left(M^TM\right)w$ algorithm and the matrix of Eq.~\ref{Eq_Metric}, the distance squared between the statepoints in the regular simplex is
\begin{equation}
	r^2_{bc}=\frac{1}{2}\sum_{i,j=1}^{d-1}\Delta{p_i}\Delta{p_j}+\frac{1}{2}\sum_{i=1}^{d-1}\left(\Delta{p_i}\right)^2.
	\label{Eq_Simplex_Distance_Initial}
\end{equation}
We can simplify this expression by using the last probabilities $p_{b_d}$ and $p_{c_d}$ of the two states' density matrices and the normalization condition, by first noting
\begin{equation}
	\Delta{p_d}\equiv{p}_{b_d}-p_{c_d}=\left(1-\sum_{i=1}^{d-1}{p}_{b_i}\right)-\left(1-\sum_{i=1}^{d-1}{p}_{c_i}\right)=-\sum_{i=1}^{d-1}\Delta{p}_i
\end{equation}
and
\begin{equation}
	\left(\Delta{p_d}\right)^2=\left(\sum_{i=1}^{d-1}\Delta{p}_i\right)\left(\sum_{j=1}^{d-1}\Delta{p}_j\right)=\sum_{i,j=1}^{d-1}\Delta{p_i}\Delta{p_j}.
\end{equation}
Substituting this expression into the first term in Eq.~\ref{Eq_Simplex_Distance_Initial} yields
\begin{equation}
	r^2_{bc}=\frac{1}{2}\left(\Delta{p_d}\right)^2+\frac{1}{2}\sum_{i=1}^{d-1}\left(\Delta{p_i}\right)^2=\frac{1}{2}\sum_{i=1}^{d}\left(\Delta{p_i}\right)^2=\frac{1}{2}||\rho_b-\rho_c||^2
	\label{Eq_Simplex_Metric}
\end{equation}
where the squared double lines indicate the (Frobenius or Hilbert-Schmidt) norm of the matrix:  the sum of the absolute value squares of each element, which in this case is just the sum of the squares of the differences in probabilities along the diagonalized matrices in the measurement simplex subspace under consideration.

\section{Decoherence Leaves}
\label{App_Decoherence}

The previous appendix dealt only with states corresponding to statepoints in the same probability simplex.  In this appendix we consider the distance between any two statepoints for two states $\hat\rho_a$ and $\hat\rho_b$.  Without loss of generality we express these states using the basis states $\ket{B_i}$ that form the measurement basis diagonalizing the $\hat\rho_b$ state:
\begin{equation}
	\hat\rho_a=\sum_{i,j=1}^da_{ij}\ket{B_i}\bra{B_j}\quad{\rm and}\quad\hat\rho_b=\sum_{i,j=1}^db_{ij}\ket{B_i}\bra{B_j}=\sum_{i=1}^db_{ii}\ket{B_i}\bra{B_i}.
	\label{Eq_rho_a_and_b}
\end{equation}
Here the $a_{ij}$ and $b_{ij}$ are the matrix elements of the representing matrices $\rho_a$ and $\rho_b$, with the last equality holding because all off-diagonal elements of $\rho_b$ are 0 in the chosen basis.

\begin{figure}
	\centering
\includegraphics[scale=.86]{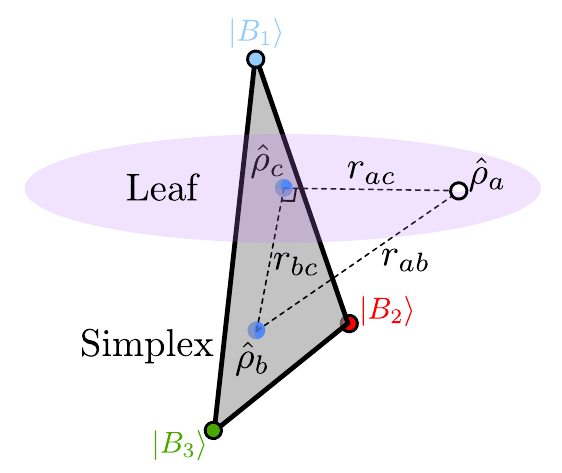}
\caption{The distance $r_{ab}$ between any $\hat\rho_a$ and $\hat\rho_b$ statepoints can be found by choosing the unique basis simplex that the $\hat\rho_b$ statepoint is on and finding the unique decoherence leaf for the simplex that contains $\hat\rho_a$.  The simplex and leaf intersect at a unique statepoint $\hat\rho_c$.  The distance $r_{bc}$ between the $\hat\rho_b$ and $\hat\rho_c$ statepoints in the simplex can be found from the probabilities of the diagonalized $\rho_b$ and $\rho_c$ matrices in the chosen basis (Appendix~\ref{App_ProbSimplex}).  The distance $r_{ac}$ between the $\hat\rho_a$ and $\hat\rho_c$ statepoints in the leaf can be found from the off-diagonal elements of their matrices (Appendix~\ref{App_Decoherence}).  The total distance  $r_{ab}$ between statepoints is then given by the Pythagorean theorem:  $r_{ab}^2=r_{ac}^2+r_{bc}^2$.}
\label{figSimLeaf}
\end{figure}

As an intermediary between these two states, it is useful to introduce the state
\begin{equation}
	\hat\rho_c\equiv\sum_{i,j=1}^da_{ij}\delta_{ij}\ket{B_i}\bra{B_i},
	\label{Eq_rho_c}
\end{equation}
 which we will see (Fig.~\ref{figSimLeaf}) is geometrically the perpendicular projection of the $\hat\rho_a$ state onto the chosen basis simplex.  The $\delta_{ij}$ Kroencker delta in the definition of this state ensures the matrix $\rho_c$ representing it in the chosen basis has off-diagonal elements equal to 0, hence corresponding to a statepoint within the basis simplex, and diagonal elements equal to the those of the $\rho_a$ matrix, meaning that $\hat\rho_a$ and $\hat\rho_c$ have identical projective value measurement probabilities when measured in this basis.

For a given basis and one diagonalized state with a statepoint in the corresponding simplex, the decoherence leaf of that state consists of all states with matching diagonal elements and measurement probabilities in that basis.  As seen from the arbitrary choice of the $\ket{B_i}$ basis and $\hat\rho_a$ state, every possible qudit state is in the decoherence leaf of one and only one diagonalized state in that basis.   Thus each choice of basis divides the entire statespace into equivalence classes of decoherence leaves, with one diagonalized state in each class.  By construction, for the chosen basis $\ket{B_i}$, the $\hat\rho_a$ and $\hat\rho_c$ statepoints are in the same decoherence leaf, which intersects the basis simplex statepoints at the single $\hat\rho_c$ statepoint at the center of the leaf (Fig.~\ref{figSimLeaf}).

It is worthwhile to consider the parametrization of a state and the corresponding location of its statepoint within its decoherence leaf. Since the matrices of all states in this space share the same diagonal elements, they are distinguished by the non-diagonal elements of these matrices.  Since the matrices are Hermitian, the independent lower-half matrix elements $a_{ij}$ with $j>i$ uniquely determine each state in a given decoherence leaf.  As there are $\frac{d\left(d-1\right)}{2}$ such independent complex matrix elements, the decoherence leaf is parametrized by $d^2-d$ real numbers.  Each leaf for a chosen basis is specified by the $d-1$ real, independent matrix diagonal elements, corresponding geometrically to the unique statepoint that is both in the measurement simplex and the leaf. Combining the parametrization needed for specifying the particular leaf and the location within it, statepoints in the full qudit statespace are located using $d^2-d+d-1=d^2-1$ real numbers.  Except for the $d=2$ qubit decoherence leaves, which are circular cross-sections of the Bloch sphere, the higher-dimensional qudit decoherence leaves are not visualizable.  However, in analogy with the single complex off-diagonal matrix element in the qubit case, which in its magnitude and argument encodes the polar coordinate position of a state within the circular cross-section in terms of its distance from and angle about the 1-d simplex diameter, the independent $\frac{d\left(d-1\right)}{2}$ off-diagonal elements of a qudit matrix each correspond to different planes in the $\left(d^2-d\right)$-dimensional leaf.  Each pair of these planes are geometrically orthogonal, but it should be emphasized that unlike two orthogonal planes in three dimensions, these planes intersect with each other at a single point (the $\hat\rho_c$ statepoint at the center of the leaf) not a line.  The location of a statepoint in the leaf is given by its polar coordinate location relative to each plane as specified by the independent off-diagonal matrix elements.

Though not even two such orthogonal planes, existing in four dimensions, are visualizable, the magnitude of each complex polar coordinate gives the distance coordinate for the $\hat\rho_a$ statepoint along that plane relative to the $\hat\rho_c$ statepoint at the center of the leaf.  Since the planes are orthogonal, the total distance $r_{ac}$ between the two statepoints is found from the sum of squares of these distances, which are the squares of the independent, off-diagonal matrix elements of $\rho_a$.  Recalling Eqs.~\ref{Eq_rho_a_and_b} and \ref{Eq_rho_c} this sum is 
\begin{equation}
	r_{ac}^2\equiv\sum_{i>j}|a_{ij}|^2=\frac{1}{2}\sum_{i\neq j}|a_{ij}|^2=\frac{1}{2}\sum_{i,j}|a_{ij}-a_{ij}\delta_{ij}|^2=\frac{1}{2}\|\rho_a-\rho_c\|^2,
	\label{Eq_Leaf_Metric}
\end{equation}
so that the matrix norm formula for this square distance between statepoints in the same decoherence leaf is identical to the Eq.~\ref{Eq_Simplex_Metric} one for the distance between statepoints in the same simplex.  As shown in Fig.~\ref{figSimLeaf}, since any line in the leaf is perpendicular to any line in the simplex, the lengths $r_{ac}$ and $r_{bc}$ form two legs of a right triangle with the square length of the hypotenuse $r_{ab}$ connecting the two arbitrary starting statepoints given geometrically by $r_{ab}^2=r_{ac}^2+r_{bc}^2$.  Using Eqs.~\ref{Eq_rho_a_and_b}, \ref{Eq_Leaf_Metric}, \ref{Eq_rho_c}, and \ref{Eq_Simplex_Metric}, this square distance can be algebraically expressed:

\begin{align}
r_{ac}^2+r_{bc}^2&=\frac12\left\|\rho_a-\rho_c\right\|^{2}+\frac12\left\|\rho_b-\rho_c\right\|^{2}\notag\\
&= \frac12\sum_{i,j}\left|a_{ij}-a_{ij}\delta_{ij}\right|^{2}
  +\frac12\sum_{i,j}\left|b_{ij}\delta_{ij}-a_{ij}\delta_{ij}\right|^{2} \notag\\
&= \frac12\sum_{i,j}\Big(
|a_{ij}|^{2}-2|a_{ij}|^{2}\delta_{ij}+|a_{ij}|^{2}\delta_{ij}^{2}
+|b_{ij}|^{2}\delta_{ij}^{2}
-\big(b_{ij}a_{ij}^{*}+b_{ij}^{*}a_{ij}\big)\delta_{ij}^{2}
+|a_{ij}|^{2}\delta_{ij}^{2}
\Big) \notag\\
&= \frac12\sum_{i,j}\left|a_{ij}-b_{ij}\delta_{ij}\right|^{2}\notag\\
&= \frac12\left\|\rho_a-\rho_b\right\|^{2}=r_{ab}^2.
\label{Eq_Gen_Metric}
\end{align}
This norm-based distance is thus justified as a the Euclidean distance between any two points in the statespace.

This distance can be put into an explicitly basis-invariant form by re-writing the norm of the general density matrix $\rho_a$ in terms of the operator $\hat\rho_a$ it represents.  From the hermitian property of any state,
\begin{equation}
	\hat\rho_a=\sum_{i,j=1}^d{a_{ij}\ket{B_i}\bra{B_j}}=\hat\rho_a^\dagger=\sum_{i,j=1}^d\left({a_{ij}\ket{B_i}\bra{B_j}}\right)^\dagger=\sum_{i,j=1}^d{a_{ij}^*\ket{B_j}\bra{B_i}}.
\end{equation}
Taking the trace of $\hat\rho_a^2$ by multiplying the first and last summation expressions and using the fact the trace is a linear operator within which bras and kets commute shows that
\begin{align}
	\Tr{\hat\rho_a^2}&=\Tr\left[\left({\sum_{i,j=1}^d{a_{ij}\ket{B_i}\bra{B_j}}}\right)\left(\sum_{k,l=1}^d{a_{kl}^*\ket{B_l}\bra{B_k}}\right)\right]\notag\\
	&=\Tr\left[{\sum_{i,j,k,l=1}^d{a_{ij}\ket{B_i}\bra{B_j}}}{a_{kl}^*\ket{B_l}\bra{B_k}}\right]=\sum_{i,j,k,l=1}^d{a_{ij}}{a_{kl}}^*\Tr\left(\ket{B_i}\bra{B_j}\ket{B_l}\bra{B_k}\right)\notag\\
	&=\sum_{i,j,k,l=1}^d{a_{ij}}{a_{kl}}^*\Tr\left(\braket{B_k|B_i}\braket{B_j|B_l}\right)=\sum_{i,j,k,l=1}^d{a_{ij}}{a_{kl}}^*\Tr\left(\delta_{ik}\delta_{jl}\right)\notag\\
	&=\sum_{i,j,k,l=1}^d{a_{ij}}{a_{kl}}^*\delta_{ik}\delta_{jl}=\sum_{i,j=1}^d{a_{ij}}{a_{ij}}^*=\sum_{i,j=1}^d|{a_{ij}}|^2=\left\|\rho_a\right\|^2\equiv\left\|\hat\rho_a\right\|^2.
	\label{Eq_Trace_Norm}
\end{align}
	We can then use this relation and the commuting of operators within the trace to write the basis-invariant form of the square distance in Eq.~\ref{Eq_Gen_Metric}:
\begin{align}
	r_{ab}^{2}
&=\frac12\left\| \hat{\rho}_a-\hat{\rho}_b\right\|^{2}
=\frac12\operatorname{Tr}\left(\hat{\rho}_a-\hat{\rho}_b\right)^{2}
=\frac12\operatorname{Tr}\left(\hat{\rho}_a^{2}+\hat{\rho}_b^{2}-\hat{\rho}_a\hat{\rho}_b-\hat{\rho}_b\hat{\rho}_a\right)\notag\\
&=\frac12\operatorname{Tr}\left(\hat{\rho}_a^{2}\right)+\frac12\,\operatorname{Tr}\left(\hat{\rho}_b^{2}\right)-\operatorname{Tr}\left(\hat{\rho}_a\hat{\rho}_b\right).
\label{Eq_Trace_Distance}
\end{align}

Being based on the absolute values of the differences of complex number elements, it is straightforward to show the Frobenius norm satisfies the Euclidean norm axioms of homogeneity, $\|c\left(\rho_a-\rho_b\right)\|=|c|\|\rho_a-\rho_b\|$ for any complex scalar $c$; positive definiteness, $\|{\rho_a-\rho_b}\|\geq0$, with the equality holding only when $\rho_a=\rho_b$; and the triangle inequality, $\|{\rho_a-\rho_c}\|+\|{\rho_c-\rho_b}\|\geq\|{\rho_a-\rho_b}\|$, with the equality holding only when the $\rho_c$ statepoint lies along the straight line between the $\rho_a$ and $\rho_b$ statepoints.

From these satisfied axioms, one can generalize to any dimension the result of Appendix~\ref{App_QubitBary} showing that the statepoint for the mixture of two qubit states lies along the line joining their statepoints, cutting it in proportion to the probability weights of the mixture.  Given the mixture $\rho=p\rho_1+\left(1-p\right)\rho_2$, then, using the homogeneity axiom,
\begin{align*}
	\|{\rho_1-\rho}\|&=\|{\rho_1-p\rho_1-\left(1-p\right)\rho_2}\|=\left(1-p\right)\|\rho_1-\rho_2\|\quad \rm{and} \\
	\|{\rho-\rho_2}\|&=\|{p\rho_1+\left(1-p\right)\rho_2-\rho_2}\|=p\|\rho_1-\rho_2\|,
\end{align*}
from which it is clear that the equality in the triangle equality axiom holds, with the $\|\rho_1-\rho_2\|$ length being cut by the $\rho$ statepoint in the ratio $\frac{1-p}{p}$.  Following the same process outlined in Appendix \ref{App_QubitBary}, the formula for the two-state mixture can be recursively used to establish the general barycentric, \textit{center of probability} algorithm for any dimensional statespace.

\section{Euclidean Statevector Distances and Angles}
\label{App_Statevector}

\begin{figure}
	\centering
\includegraphics[scale=.46]{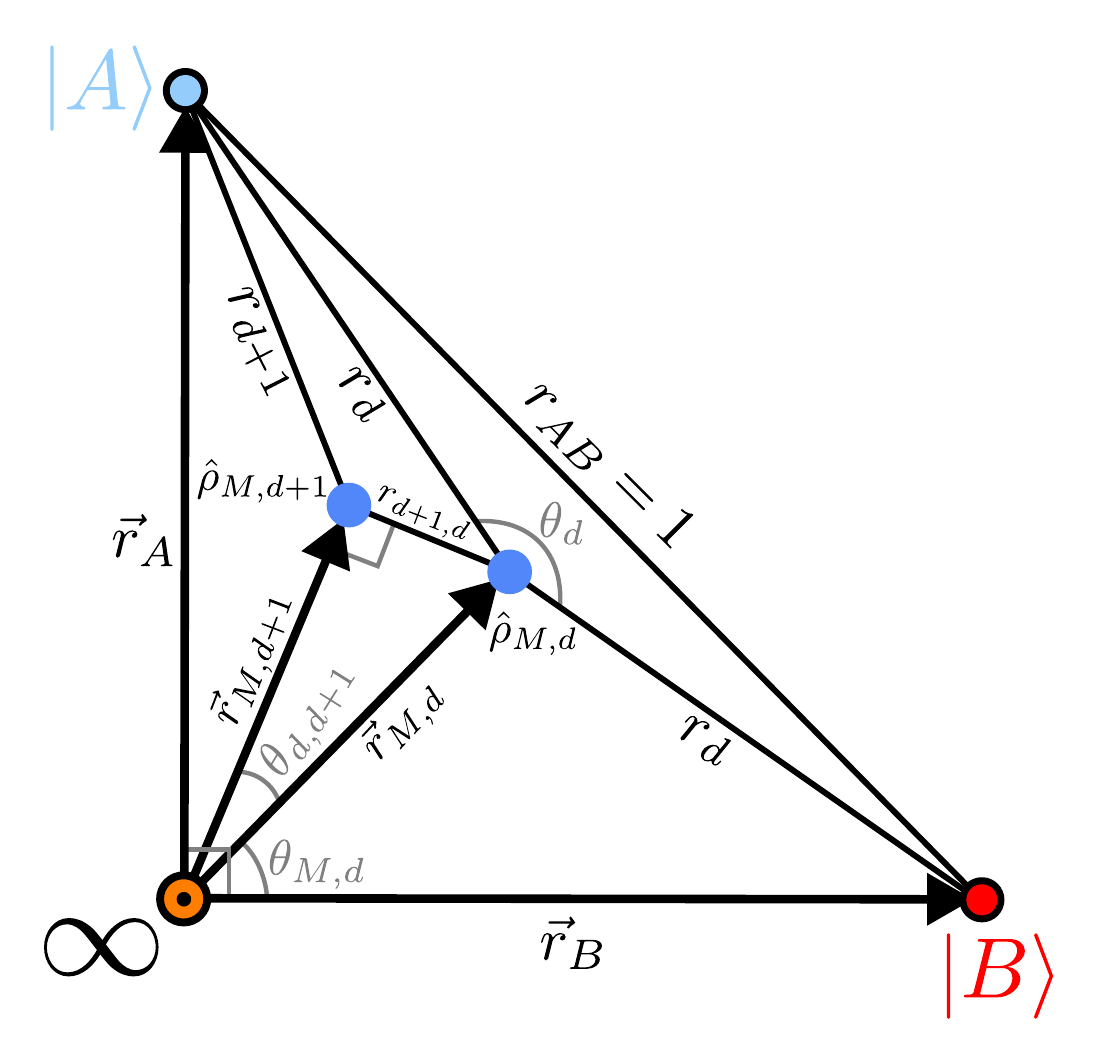}
\caption{Two orthogonal pure statepoints, $\ket{A}$ and $\ket{B}$; two successive maximally mixed statepoints, $\hat\rho_{M,d}$ and $\hat\rho_{M,d+1}$, that are mixtures of them; the maximally mixed statepoint, $\hat\rho_\infty\equiv\hat\rho_{M,\infty}$; the lines connecting them; and the angles between these lines.}
\label{figTriangles}
\end{figure}

The quoted results of Section~\ref{statevectors} involve the calculation of lengths and angles found in Fig.~\ref{figInfOrigin}(c), which for ease of reference is reproduced as Fig.~\ref{figTriangles}.  The calculations are useful exercises for students learning the trace and density-operator formalism as they can be done using the trace form of the general metric between two statepoints,
\begin{align}
	r_{bc}^{2}=\frac12\,\operatorname{Tr}\!\left(\hat{\rho}_b^{2}\right)+\frac12\,\operatorname{Tr}\!\left(\hat{\rho}_c^{2}\right)-\operatorname{Tr}\!\left(\hat{\rho}_b\hat{\rho}_c\right),
\label{Eq_Trace_Distance_2}
\end{align}
and the understanding that bras, kets, and operators commute within the trace operation.  As a preliminary, it is useful to recall that the normalization of any quantum state $\hat\rho$ requires that $\operatorname{Tr}\left(\hat\rho\right)=1$; that for any pure state $\rho_A\equiv\ket{A}\bra{A}$,
\begin{equation}
	\operatorname{Tr}\left({\hat\rho_A}^2\right)=\operatorname{Tr}\left(\left(\ket{A}\bra{A}\right)\left(\ket{A}\bra{A}\right)\right)=\operatorname{Tr}\left(\braket{A|A}\braket{A|A}\right)=\operatorname{Tr}\left(1\right)=1;
	\label{Eq_PureTrace}
\end{equation}
and that for a maximally mixed state of dimension $d$ expressed in terms of an orthonormal basis, $\hat\rho_{M,d}=\sum_{i=1}^d\frac{1}{d}\ket{B_i}\bra{B_i}$,
\begin{align}	
	\operatorname{Tr}\left({\hat\rho_{M,d}}^2\right)&=\operatorname{Tr}\left(\sum_{i=1}^d\frac{1}{d}\ket{B_i}\bra{B_i}\sum_{j=1}^d\frac{1}{d}\ket{B_j}\bra{B_j}\right)=\operatorname{Tr}\left(\frac{1}{d^2}\sum_{i,j=1}^d\braket{B_i|B_j}\braket{B_j|B_i}\right)\notag\\
	&=\frac{1}{d^2}\operatorname{Tr}\left(\sum_{i,j=1}^d\delta_{i,j}^2\right)=\frac{1}{d^2}\operatorname{Tr}\left(\sum_{i=1}^d\delta_{i,i}^2\right)=\frac{1}{d^2}d=\frac{1}{d}.
\label{Eq_MaxMixedTrace}
\end{align}

Using Eq.~\ref{Eq_PureTrace}, the square distance between any two orthonormal pure states $\ket{A}$ and $\ket{B}$ is
\begin{align}
r_{AB}^{2}
&= \frac12\operatorname{Tr}\left(\hat{\rho}_A^{2}\right)
 + \frac12\operatorname{Tr}\left(\hat{\rho}_B^{2}\right)
 - \operatorname{Tr}\left(\hat{\rho}_A\hat{\rho}_B\right)
= \frac12 + \frac12 - \operatorname{Tr}\left(|A\rangle\langle A|B\rangle\langle B|\right)\notag \\
&= 1 - \operatorname{Tr}\!\left(\langle B|A\rangle\,\langle A|B\rangle\right) = 1-\operatorname{Tr}(0)=1,
\end{align}
so that $r_{AB}=1$.

The distance between a general state $\hat\rho_a$, expressible in any finite-dimensional orthonormal basis as $\hat\rho_a=\sum_{i,j=1}^da_{ij}\ket{B_i}\bra{B_j}$, and the maximally mixed state $\hat\rho_{M,d}$ at the center of this basis simplex is
\begin{align}
r_{a,{\left(M,d\right)}}^{2}
&= \frac12\operatorname{Tr}\left(\hat{\rho}_a^{2}\right)
 + \frac12\operatorname{Tr}\left(\hat{\rho}_{M,d}^{2}\right)
 - \operatorname{Tr}\!\left(\hat{\rho}_a\hat{\rho}_{M,d}\right) \notag\\
&= \frac12\operatorname{Tr}\left(\hat{\rho}_a^2\right)
 + \frac12\left(\frac{1}{d}\right)
 - \operatorname{Tr}\left(\sum_{i,j=1}^da_{ij}\ket{B_i}\bra{B_j}\sum_{k=1}^d{\frac{1}{d}}\ket{B_k}\bra{B_k}\right) \notag\\
&= \frac12\operatorname{Tr}\left(\hat{\rho}_a^2\right)
 + \frac{1}{2d}
 - \operatorname{Tr}\left(\frac{1}{d}\sum_{i,j,k=1}^da_{ij}\ket{B_i}\braket{B_j|B_k}\bra{B_k}\right) \notag\\
&= \frac12\operatorname{Tr}\left(\hat{\rho}_a^2\right)
 + \frac{1}{2d}
 - \frac{1}{d}\operatorname{Tr}\left(\sum_{i,j,k=1}^da_{ij}\ket{B_i}\delta_{j,k}\bra{B_k}\right) \notag\\
&= \frac12\operatorname{Tr}\left(\hat{\rho}_a\right)
 + \frac{1}{2d}
 - \frac{1}{d}\operatorname{Tr}\left(\sum_{i,j=1}^da_{ij}\ket{B_i}\bra{B_j}\right) \notag\\
&= \frac12\,\operatorname{Tr}\!\left(\hat{\rho}_a^{2}\right)
 + \frac{1}{2d}
 - \frac{1}{d}\,\operatorname{Tr}\!\left(\hat{\rho}_a\right) = \frac12\,\operatorname{Tr}\!\left(\hat{\rho}_a^{2}\right)
 + \frac{1}{2d}-\frac{1}{d}\left(1\right)= \frac12\,\operatorname{Tr}\!\left(\hat{\rho}_a^{2}\right)-\frac{1}{2d},
 \label{Eq_MaxGen}
\end{align}
where we have used the normalization condition on the trace of any state in the last steps.  From the Eq.~\ref{Eq_MaxGen} result and the unit trace of the square of pure states of Eq.~\ref{Eq_PureTrace}, the distance between a maximally mixed state and a pure state in its statespace shown in of Fig.~\ref{figTriangles} is $r_d=\sqrt{\left(\frac12-\frac{1}{2d}\right)}$.  From Eq.~\ref{Eq_MaxGen} and the Eq.~\ref{Eq_MaxMixedTrace} trace of a maximally mixed state one can find that the distance between a maximally mixed state of dimension $d$ and one of lower dimension $d'$ within its statespace is $r_{d,d'}=\sqrt{\left(\frac{1}{2d'}-\frac{1}{2d}\right)}$. In particular, for the successive maximally mixed states $\hat\rho_{M,d}$ and $\hat\rho_{M,d+1}$ of Fig.~\ref{figTriangles}, the statepoint distance is $r_{d+1,d}=\frac{1}{\sqrt{2d\left(d+1\right)}}$.  Finally, from the Eq.~\ref{Eq_MaxGen} result in the $d\rightarrow\infty$ limit, the length of the statevectors drawn from the $\hat\rho_{M,\infty}\equiv\hat\rho_{\infty}$ origin to any state $\hat\rho_a$ is given by $r_a=\sqrt{\frac12\rm{Tr}\left(\hat\rho_a^2\right)}$.  In particular, the lengths of statevectors to pure states, such as $\ket{A}$ and $\ket{B}$, are $r_A=r_B=\frac{1}{\sqrt{2}}$, and the lengths to maximally mixed states, such as $\hat\rho_{M,d}$ and $\hat\rho_{M,d+1}$, are $r_{M,d}=\frac{1}{\sqrt{2d}}$ and $r_{M,d+1}=\frac{1}{\sqrt{2\left(d+1\right)}}$.

With all the distances of Fig.~\ref{figTriangles} calculated, we proceed to find the marked angles.  Since 
\begin{equation}
	r_A^2+r_B^2=\frac12+\frac12=1=r_{AB}^2
\end{equation}
we see that the two statevectors for any pair of orthogonal pure states always form the legs of a right triangle.  Similarly, the fact that
\begin{equation}
	r_{M,d+1}^2+r_{d+1,d}^2=\frac{1}{2\left(d+1\right)}+\frac{1}{2d\left(d+1\right)}=\frac{d+1}{2d\left(d+1\right)}=\frac{1}{2d}=r_{M,d}^2
\end{equation}
shows that the triangle made by connecting the endpoints of successive maximally mixed statevectors $\vec{r}_{M,d}$ and $\vec{r}_{M,d+1}$ is also a right triangle.  The angle $\theta_{d,d+1}$ between these two statevectors is then found trigonometrically:
\begin{align}
	\cos{\theta_{d,d+1}}&=\frac{r_{M,d+1}}{r_{M,d}}=\sqrt{\frac{2d}{2\left(d+1\right)}},\notag\\
	\implies\theta_{d,d+1}&=\cos^{-1}\left(\sqrt{\frac{d}{d+1}}\right).
\end{align}

The isosceles triangle made by connecting the center of a $d-1$ regular simplex (a $\hat\rho_{M,d}$ statepoint) to two of its vertices joined by a simplex edge (two orthogonal pure statepoints $\ket{A}$ and $\ket{B}$ in its mixture) has an angle $\theta_d$ subtending the $r_{AB}$ unit-length side of the triangle connecting the pure statepoints.  By the law of cosines,
\begin{align}
	&\quad\quad\quad r_d^2+r_d^2-2r_d^2\cos{\theta_d}=2r_d^2\left(1-\cos{\theta_d}\right)= r_{AB}^2,\notag\\
	&\implies\theta_d=\cos^{-1}\left(1-\frac{r_{AB}^2}{2r_d^2}\right)=\cos^{-1}\left(1-\frac{1}{1-\frac{1}{d}}\right)=\cos^{-1}\left(\frac{1}{1-d}\right).
\end{align}   
The $\theta_{M,d}$ angle between a maximally mixed statevector $\vec{r}_{M,d}$ and any statevector $\vec{r}_{B}$ of a pure state in the mixture can similarly be found by completing the triangle connecting the $\hat\rho_{M,d}$ and $\hat\rho_{B}$ statepoints at the ends of these vectors with the $r_d$ line and using the law of cosines:
\begin{align}
	r_{M,d}^2+r_B^2-2r_{M,d}r_B\cos{\theta_{M,d}}&=r_d^2,\notag\\
	\implies\frac{1}{2d}+\frac12-2\frac{1}{\sqrt{2d}}\frac{1}{\sqrt{2}}\cos{\theta_{M,d}}&=\frac12-\frac{1}{2d},\notag\\
	\implies\theta_{M,d}=\cos^{-1}\left({\frac{1}{\sqrt{d}}}\right).&
\end{align}

Finally, we note that the trace of products of pairs of hermitian operators and matrices that is present in the general metric and many other algorithms is equivalent to the Frobenius product of operators and matrices, which is defined for two matrices $\rho_a$ and $\rho_b$ with components $a_{ij}$ and $b_{ij}$ as
\begin{equation}
	\braket{\rho_a,\rho_b}_F=\sum_{i,j}a_{i,j}^*b_{i,j}.
\end{equation}
This product is the natural extension of the inner product of pure states familiar to starting quantum students, where corresponding entries of complex-conjugated row vectors and complex column vectors are multiplied and summed, and thus emphasizes both the extension of that formalism and its close relation to the geometrical dot product of real, Euclidean statevectors.  We have elected to use the trace formalism in this work as this operation is both widely used in the context of mixed states and one that starting quantum students have little practice with.  One of the central virtues of the trace operation that has been emphasized in these proofs is the commutativity of bras, kets, and operators within the trace. After the challenge of acclimating to the non-commuting algebra of quantum mechanics, this restoration of commutativity can be a welcome relief as reflected in this limerick:

\begin{quotation}
\noindent\textit{The trace is a wonderful place.\\
For in its warm embrace,\\
With ease and grace you will compute\\
For kets and bras can now commute\\
In that wonderful place called the trace.}\\
\end{quotation}

\end{document}